\documentclass[11pt,twoside]{article}

\usepackage{pslatex}    
\usepackage{amsfonts}
\usepackage{amsmath}
\usepackage{amsthm}
\usepackage{amscd}
\usepackage{cite}
\usepackage{eucal}
\usepackage{epsf}
\usepackage{epsfig}
\usepackage{a4}
\usepackage{booktabs}

\newcommand{\tb}{\mathbf{t}}
\renewcommand{\sb}{\mathbf{s}}
\newcommand{\bR}{\mathbb{R}}

\def\beq{\begin{equation}}
\def\eeq{\end{equation}}
\def\bea{\begin{eqnarray}}
\def\eea{\end{eqnarray}}

\def\beann{\begin{eqnarray*}}
\def\eeann{\end{eqnarray*}}

\let\a=\alpha \let\be=\beta \let\g=\gamma \let\de=\delta
\let\e=\varepsilon \let\z=\zeta \let\h=\eta 
\let\eps=\epsilon
 \let\k=\kappa \let\la=\lambda \let\m=\mu
\let\n=\nu \let\x=\xi \let\p=\pi \let\r=\rho \let\s=\sigma
\let\om=\omega \let\ps=\psi
\let\ph=\varphi \let\Ph=\phi \let\PH=\Phi \let\Ps=\Psi
\let\Om=\Omega  
\let\La=\Lambda \let\G=\Gamma \let\D=\Delta

\let\qd=\quad  

\def\epp{\, .}
\def\epc{\, ,}

\def\tst#1{{\textstyle #1}}
\def\dst#1{{\displaystyle #1}}

\theoremstyle{plain}

\newtheorem*{corollary*}{Corollary}

\newtheorem*{conjecture*}{Conjecture}

\theoremstyle{definition}

\def\2{\frac{1}{2}} \def\4{\frac{1}{4}}

\def\6{\partial}

\def\+{\dagger}

\def\<{\langle} \def\>{\rangle}

\def\CO{{\cal O}}

\def\i{{\rm i}}

\def\rd{{\rm d}}
\def\re{{\rm e}}

\DeclareMathOperator{\sh}{sh}
\DeclareMathOperator{\ch}{ch}

\DeclareMathOperator{\cth}{cth}

\DeclareMathOperator{\tr}{tr}

\DeclareMathOperator{\End}{End}

\DeclareMathOperator{\res}{res}

\def\Re{{\rm Re\,}} \def\Im{{\rm Im\,}}

\def\bv{\mathbf{b}}
\def\cv{\mathbf{c}}

\def\ev{\mathbf{e}}

\def\tv{\mathbf{t}}
\def\uv{\mathbf{u}}
\def\vv{\mathbf{v}}

\def\Uv{\mathbf{U}}
\def\Vv{\mathbf{V}}
\def\Wv{\mathbf{W}}

\def\fa{\mathfrak{a}}

\def\faq{\overline{\mathfrak{a}}}

\def\fb{\mathfrak{b}}
\def\fbq{\overline{\mathfrak{b}}}

\renewcommand{\appendix}{%
   \renewcommand{\section}{
        \secdef\Appendix\sAppendix}%
   \setcounter{section}{0}%
   \renewcommand{\thesection}{\Alph{section}}%
   \renewcommand{\theequation}{\thesection.\arabic{equation}}%
}

\newcommand{\Appendix}[2][?]{%
     \refstepcounter{section}%
     \setcounter{equation}{0}%
     \addcontentsline{toc}{appendix}%
          {\protect\numberline{\appendixname~\thesection} #1}%
     \vspace{\baselineskip}%
     {\noindent\large\bfseries\appendixname\ \thesection: #2\par}%
     \sectionmark{#1}\vspace{\baselineskip}}

\newcommand{\sAppendix}[1]{%
     {\noindent\large\bfseries\appendixname\:: #1\par}%
     \sectionmark{#1}\vspace{\baselineskip}}

\renewcommand{\tilde}{\widetilde}
\renewcommand{\arraystretch}{1.1}
\newcommand{\slt}{\mathfrak{sl}_2}

\begin{document}

\thispagestyle{empty}

\begin{center}

{\Large {\bf On the physical part of the factorized correlation functions
of the XXZ chain\\}}

\vspace{7mm}

{\large
Herman Boos\footnote{e-mail: boos@physik.uni-wuppertal.de} and
Frank G\"{o}hmann\footnote{e-mail: goehmann@physik.uni-wuppertal.de}}\\

\vspace{5mm}

{Fachbereich C -- Physik, Bergische Universit\"at Wuppertal,\\
42097 Wuppertal, Germany\\}

\vspace{20mm}

{\large {\bf Abstract}}

\end{center}

\begin{list}{}{\addtolength{\rightmargin}{9mm}
               \addtolength{\topsep}{-5mm}}
\item
It was recently shown by Jimbo, Miwa and Smirnov that the correlation
functions of a generalized XXZ chain associated with an inhomogeneous
six-vertex model with disorder parameter $\alpha$ and with arbitrary
inhomogeneities on the horizontal lines factorize and can all be expressed
in terms of only two functions $\rho$ and $\omega$. Here we approach the
description of the same correlation functions and, in particular, of the
function $\omega$ from a different direction. We start from
a novel multiple integral representation for the density matrix of a finite
chain segment of length $m$ in the presence of a disorder field $\alpha$.
We explicitly factorize the integrals for $m=2$. Based on this we
present an alternative description of the function $\omega$ in terms
of the solutions of certain linear and nonlinear integral equations.
We then prove directly that the two definitions of $\omega$ describe
the same function. The definition in the work of Jimbo, Miwa and Smirnov
was crucial for the proof of the factorization. The definition given
here together with the known description of $\rho$ in terms of the
solutions of nonlinear integral equations is useful for performing e.g.\
the Trotter limit in the finite temperature case, or for obtaining
numerical results for the correlation functions at short distances.
We also address the issue of the construction of an exponential form
of the density matrix for finite $\alpha$.
\\[2ex]
\end{list}

\clearpage

\section{Introduction}
In recent years significant progress has been achieved in the
understanding of the mathematical structure of the correlation functions
of the XXZ model and related integrable models. 
First of all the ground state correlation functions were studied.
They are completely defined through the quantum-mechanical density
matrix. An explicit expression for the density matrix of a finite
subchain of the infinite XXZ chain in the massive regime was first
obtained by Jimbo, Miki, Miwa and Nakayashiki \cite{JMMN92}. They
expressed the elements of the density matrix in terms of multiple
integrals. Subsequently, extensions of their formulae to the massless
regime and to a non-vanishing longitudinal magnetic field were obtained
in \cite{JiMi96,KMT99b}.

Then it was realized that the multiple integrals can be factorized
\cite{BoKo01} and that, utilizing the so-called reduced
Knizhnik-Zamolodchikov equation, the factorized integrals can be
written in a compact exponential form \cite{BJMST04a,BJMST04b}. The
latter allows one to distinguish between an algebraic part and a physical
part. The physical part is defined by a small number of transcendental 
functions, fixed by the one-point-correlators and by the two-point
neighbour correlators which depend on the physical parameters like
anisotropy, temperature, length of the chain, magnetic field, boundary
conditions etc. The algebraic part is related to the representation
theory of the symmetry algebra behind the model, namely the quantum
group $U_q(\widehat{\slt})$ in case of the XXZ chain.

In \cite{BJMST06b} it was observed that the formula for the correlation
functions looks nicer if the XXZ chain is regularized by introducing an
additional parameter, the disorder field~$\a$. With this new parameter 
it was possible to express the density matrix in terms of special
fermionic annihilation operators $\bv$ and $\cv$ acting not on states
of the spin chain, but on the space of (quasi-) local operators on
these states. The annihilation operators appeared to be responsible for
the algebraic part. The physical part was represented by a transcendental
function $\om$ determined by a single integral. In \cite{BJMST08app}
the dual fermionic creation operators $\bv^*$, $\cv^*$ and a bosonic
creation operator $\tv^*$ were constructed. These operators together
generate a special basis of the space of quasi-local operators. Since
$\bv^*$, $\cv^*$ are Fermi operators, Wick's theorem applies and
expectation values of products of $\bv^*$, $\cv^*$ and $\tv^*$ in an
appropriately defined vacuum state can be written as determinants, very
much as in the case of free fermions.

Thermodynamic properties of integrable lattice models can be
studied within the Suzuki-Trotter formalism by considering
an auxiliary lattice with staggering in the so-called
Trotter direction \cite{Suzuki85}. The temperature appears as a result
of a special limit when the extension of the lattice in
Trotter direction becomes infinite. Physical quantities are
expressed in an efficient way through the solution to certain
non-linear integral equations \cite{Kluemper92}. A detailed discussion
of this issue and further references can, for instance, be found in
the book \cite{thebook}. 

In papers \cite{GKS04a,GKS05} the Suzuki-Trotter formalism was used
in order to generalize the multiple integrals to finite temperature.
Then their factorization was probed for several examples of correlation
functions, first for the XXX chain \cite{BGKS06} and later for the
XXZ chain \cite{BGKS07,BDGKSW08}. Also a conjecture was formulated
stating that the above mentioned exponential form is valid with the
same fermionic operators (at least as long as they act on spin reversal
invariant products of local operators) as for the ground state and
two functions $\om$, $\om'$ obtained from an $\a$-dependent function
in the limit $\a\rightarrow 0$. 

Unfortunately, the formulae of \cite{BGKS07,BDGKSW08} worked only in
this limit. The generalization to generic $\a$ stayed obscure. One of
the purposes of the present work is to add to the clarification of this
point, starting from a proper multiple integral representation with
disorder parameter $\a$. Here, as we had to learn \cite{Kitanine08up},
the crucial point is that the `Cauchy extraction trick', invented in
\cite{IKMT99} and described in more detail in \cite{KKMST09a}, can be
applied in the finite temperature case and also in the more general
situation of a finite lattice with inhomogeneities in Trotter direction.

Important new insight came from a recent paper \cite{JMS08pp} by Jimbo,
Miwa and Smirnov, where they suggested a purely algebraic
approach to the problem of calculating the static correlation
functions of the XXZ model.
The key idea of \cite{JMS08pp} is to evaluate a linear functional
related to the partition function within the fermionic basis constructed
in \cite{BJMST08app}.
The authors of \cite{JMS08pp} work with a finite lattice, inhomogeneous
in Trotter direction. In this situation they suggest a new and
surprising construction of the function $\om$ depending on a magnetic
field and on the disorder parameter $\a$.

In the present paper we discuss the relation of the work by Jimbo,
Miwa and Smirnov to the approach using non-linear integral equations
which at the moment seems more appropriate e.g.\ for taking the
Trotter limit (which was omitted in \cite{JMS08pp}). In particular,
we present an alternative description of the function $\om$ starting
from the multiple integral and using the explicit factorization
of the density matrix for two neighbouring lattice sites. We then
give a direct proof that our expression, though looking rather
different than that in \cite{JMS08pp}, in fact describes the same 
function.

An inhomogeneous lattice in Trotter direction is very general and
leaves many different options for the realization of physical correlation
functions. Here we shall concentrate on two of them, the correlation
functions of the infinite XXZ chain at finite temperature and magnetic
field (temperature case) and the ground state correlation functions
of a finite chain with twisted periodic boundary conditions (finite
length case). Both cases can be treated to a very large extend
simultaneously. They are only distinct in that a different distribution
of inhomogeneity parameters is required and in that for the finite
temperature case the Trotter limit has to be performed. Note that
instead of the XXZ Hamiltonian we could consider combinations of
conserved quantities obtained from the transfer matrix of the six-vertex
model within the formalism of non-linear integral equations. For the
bulk thermodynamic properties this issue was recently studied in
\cite{TrKl07}.

The paper is organized as follows. In the next section we define our
basic objects and recall some of their properties. In the third
section we show the multiple integral formula for the elements of
the ($\a$-twisted) density matrix for a sub-chain of length~$m$.
In section four we consider the simplest case $m = 1$. The fifth
section is devoted to applying the factorization technique to the
double integrals for $m = 2$. In section~\ref{sec:om} we introduce
the function $\om$. We discuss its properties and the relation to 
its realization by Jimbo, Miwa and Smirnov. The content of
section~\ref{sec:top} is some preliminary work on the construction
of an operator $\tv$, dual to the creation operator $\tv^*$, which
should appear in the construction of an exponential form for finite
temperature and finite disorder parameter. In the appendices we
provide a derivation of the multiple integral formulae, we discuss
the limit $\a \rightarrow 0$, and we compare with the results of the
papers \cite{BGKS07,BDGKSW08}. 

\section{Density matrix and correlation functions}
The XXZ quantum spin chain is defined by the Hamiltonian
\begin{equation} \label{xxzham}
     H_N (\k) = J \sum_{j=1}^N \bigl( \s_{j-1}^x \s_j^x
           + \s_{j-1}^y \s_j^y + \D (\s_{j-1}^z \s_j^z - 1) \bigr) \epc
\end{equation}
written here in terms of the Pauli matrices $\s^x = e_-^+ + e_+^-$,
$\s^y = \i (e_-^+ - e_+^-)$, $\s^z = e_+^+ - e_-^-$ (where the $e^\a_\be$
are the elements of the gl(2) standard basis). The two real parameters
$J$ and $\D$ control the ground state phase diagram of the model.
For simplicity of notation we shall restrict ourselves in the following
to the critical phase $J > 0$, $|\D| < 1$. Note, however, that the
results of this work can be easily extended to the off-critical
antiferromagnetic phase $\D > 1$. We shall also assume without further
mentioning that the number of lattice sites $N$ is even.

To fully specify $H_N (\k)$ we have to define the boundary conditions.
We shall consider twisted periodic boundary conditions, when we
are dealing with the ground state of the finite chain. Then $H_N (\k)$
depends on an additional parameter $\k$ through
\begin{equation} \label{twistbound}
     \begin{pmatrix} {e_0}_+^+ & {e_0}_-^+ \\
                     {e_0}_+^- & {e_0}_-^- \end{pmatrix} =
     q^{- \k \s^z} \begin{pmatrix} {e_N}_+^+ & {e_N}_-^+ \\
                      {e_N}_+^- & {e_N}_-^-
		   \end{pmatrix} q^{\k \s^z} \epp
\end{equation}
Here $q$ is related to $\D$ as $\D = (q + q^{-1})/2$. For the finite
temperature case we shall assume periodic boundary conditions for
the Hamiltonian. Nevertheless the same parameter $\k$ will appear
in that case as a twist parameter of the quantum transfer matrix, having
then a rather different physical meaning as an external magnetic field
coupling to the spins by a Zeeman term. We shall elaborate on this
below.

The integrable structure behind the Hamiltonian (\ref{xxzham}) is
generated by the trigonometric $R$-matrix of the six-vertex model
\cite{Babook},
\begin{align} \label{rxxz}
     R(\la) & = \begin{pmatrix}
                    1 & 0 & 0 & 0 \\
		    0 & b(\la) & c(\la) & 0 \\
		    0 & c(\la) & b(\la) & 0 \\
		    0 & 0 & 0 & 1
		\end{pmatrix} \epc \\[2ex]
     b(\la) & = \frac{\sh(\la)}{\sh(\la + \h)} \epc \qd
     c(\la) = \frac{\sh(\h)}{\sh(\la + \h)} \epc \label{defbc}
\end{align}
acting on ${\mathbb C}^2 \otimes {\mathbb C}^2$. As presented here it
satisfies the Yang-Baxter equation in additive form. To facilitate the
comparison with \cite{BJMST08app,JMS08pp}, where the multiplicative form
was preferred, we set $q = \re^\h$ and $\z = \re^\la$. Then for arbitrary
complex inhomogeneity parameters $\be_j$, $j = 1, \dots, N$, the definition
\begin{equation} \label{defmono}
     T_a (\z) = R_{a, N} (\la - \be_N) \dots R_{a, 1} (\la - \be_1)
\end{equation}
of the monodromy matrix makes sense, where, as usual, the indices
 $1, \dots,N$ refer to the spin chain while $a$ refers to an additional
site defining the so-called auxiliary space. We also set $T_a (\z, \k) =
T_a (\z) q^{\k \s^z_a}$ and introduce the twisted transfer matrix
\begin{equation}
     t(\z, \k) = \tr_a \bigl( T_a (\z, \k) \bigr) \epp
\end{equation}

In \cite{JMS08pp} a six vertex-model with $N$ horizontal rows and
an arbitrary distribution of the inhomogeneities $\tau_j = \re^{\be_j}$
on these rows was considered. Here we would like to point out that two
specific distributions are of particular interest in physical
applications. Moreover, in both cases the special functions that enter
the representations of the transfer matrix eigenvalues and correlation
functions have nice descriptions in terms of solutions of linear and
non-linear integral equations.

The first case relates to the ground state of the Hamiltonian
(\ref{xxzham}). We call it the finite length case. In this case we choose
\begin{equation} \label{tdistr}
     \be_j = \h/2 \epc \qd j = 1, \dots, N \epp
\end{equation}
Then
\begin{equation} \label{hamfromt}
     H_N (\k) = 2 J \sh(\h) \,
           \6_\la \ln \bigl( t^{-1}(q^\2,\k) \, t(\z,\k) \bigr)%
	   \big|_{\la = \h/2} \epc
\end{equation}
with twisted boundary conditions (\ref{twistbound}) if we identify
$\D = \ch (\h)$. The critical regime $|\D| < 1$ corresponds to purely
imaginary $\h = \i \g$, $\g \in [0, \p)$. In this case the physical twist
angle or flux $\PH \in [0,2\p)$ is $\PH = - \k \g$, whence $\k$ should
be real.  If we stick to the vertex model picture of \cite{JMS08pp},
then $t(\z,\k)$ is the vertical or column-to-column transfer matrix in
this case.

The second case is determined by an alternating choice
\begin{equation} \label{qtmdistr}
     \be_j = \begin{cases} \be_{2j-1} = \h - \frac{\be}{N} \\
                           \be_{2j} = \frac{\be}{N}
             \end{cases} \epc \qd j = 1, \dots, N/2 \epc
\end{equation}
of inhomogeneity parameters. This case will be called the finite
temperature case as it relates to the quantum transfer matrix, whose
monodromy matrix is
\begin{multline}
     T^{QTM}_a (\z) = \\ R_{a, N} (\la - \be/N)
                      R_{N-1, a}^{t_1} (- \be/N - \la) \dots
                      R_{a, 2} (\la - \be/N)
                      R_{1, a}^{t_1} (- \be/N - \la) \epp
\end{multline}
Here the superscript `$t_1$' indicates transposition with respect to the
first space. In fact, setting $Y = \prod_{j=1}^{N/2} \s_{2j-1}^y$ and
using the crossing symmetry
\begin{equation}
     \s_j^y R_{a, j} (\la - \h) \s_j^y = b(\la - \h) R_{j, a}^{t_1} (- \la)
\end{equation}
of the $R$-matrix we find that
\begin{equation} \label{ttqtm}
     T^{QTM}_a (\z) = Y T_a (\z) Y
                      \prod_{j=1}^{N/2} \frac{1}{b(\la - \be_{2j-1})} \epp
\end{equation}
The quantum transfer matrix is by definition
\begin{equation}
     t^{QTM} (\z,\k) = \tr_a \bigl( T^{QTM}_a (\z, \k) \bigr) \epc
\end{equation}
where $T^{QTM}_a (\z, \k) = T^{QTM}_a (\z) q^{\k \s^z_a}$.

Again, within the vertex model picture, $t^{QTM} (\z,\k)$, or $t (\z,\k)$
with the choice (\ref{qtmdistr}) of inhomogeneity parameter, corresponds
to the vertical transfer matrix. There is an important difference,
though, that has been explained at several occasions \cite{Kluemper92,%
GKS04a}. In the finite length case the Hamiltonian can be derived
from the vertical transfer matrix. In particular, the vertical
transfer matrix and the Hamiltonian (\ref{xxzham}) have the same
eigenstates. In the finite temperature case, on the other hand,
with a lattice which is homogeneous in horizontal direction, say, the
Hamiltonian is related to the horizontal transfer matrix with purely
periodic boundary conditions. It is then also periodic and will be
denoted $H_L (0)$, where $L$ is the horizontal extension of the lattice.
In this case the vertical transfer matrix eigenstates are different
from those of the Hamiltonian. In particular, the eigenstate with
the largest modulus determines the state of thermodynamic equilibrium
in the thermodynamic limit, i.e.\ the free energy of the XXZ chain
and all its static correlation functions \cite{GKS04a}. Also the
physical interpretation of the parameter $\k$ is rather different in
this case. It corresponds to a magnetic field coupling to the spin chain
through a Zeeman term (see e.g.\ \cite{GKS04a}).

Using a lattice of finite extension $L$ in horizontal direction we can
express the partition function of the homogeneous XXZ chain of length $L$
as
\begin{equation} \label{zustand}
     Z_L = \tr_{1, \dots, L} \re^{- \be H_L (0) + h S_{[1,L]}/T}
         = \lim_{N \rightarrow \infty} \tr_{1, \dots, N}
	   \bigl( t^{QTM} (1,h/(2 \h T)) \bigr)^L \epp
\end{equation}
Here $T$ is the temperature and $h$ is a longitudinal magnetic field.
$\be$ must be chosen as $\be = 2J \sh(\h)/T$. Furthermore
\begin{equation}
     S_{[1,L]} = \tst{\2} \sum_{j=1}^L \s_j^z
\end{equation}
is the conserved $z$-component of the total spin. Equation (\ref{zustand})
becomes efficient in the thermodynamic limit $L \rightarrow \infty$,
since then a single eigenvalue $\La^{QTM} (1, \k)$ of $t^{QTM} (1,\k)$
of largest modulus dominates the large-$L$ asymptotics of $Z_L$ in the
Trotter limit $N \rightarrow \infty$. We shall refer to this eigenvalue
as the dominant one.

We would like to remark that in our understanding the quantum transfer
matrix is, in general, more appropriate for studying integrable spin
models on the infinite lattice than the usual transfer matrix. In
general there is no crossing symmetry, and the quantum transfer matrix
and the usual transfer matrix are not related by a similarity
transformation like in (\ref{ttqtm}). Also within the quantum transfer
matrix formulation the density matrix directly takes its `natural form' in
terms of monodromy matrix elements (see below). No solution of a quantum
inverse problem as in \cite{KMT99b} is required. In our particular case
we do have the crossing symmetry, and the quantum transfer matrix
and the usual transfer matrix with staggered inhomogeneities
(\ref{qtmdistr}) give an equivalent description of the density matrix
(see below). Still, the largest eigenvalue of $t(\z,\k)$ with the
distribution (\ref{qtmdistr}) of inhomogeneities diverges in the Trotter
limit as can be seen from (\ref{ttqtm}).

Let us come back to the situation of arbitrarily distributed
inhomogeneity parameters $\be_j$. Following \cite{JMS08pp} we shall
assume that for a certain spectral parameter $\z_0$ and any
$\k \in {\mathbb C}$ the transfer matrix $t(\z_0,\k)$ has a unique
eigenvector $|\k\>$ with eigenvalue $\La (\z_0, \k)$ of largest modulus.
This is certainly true for the two special cases considered above.
In the finite length case $\z_0 = q^{1/2}$, while $\z_0 = 1$ in the
finite temperature case. We fix a set of `vertical inhomogeneity
parameters' $\n_1, \dots, \n_m$ and set $\x_j = \re^{\n_j}$. Then we
can define the object of our main interest, the density matrix
with matrix elements
\begin{equation} \label{defdens}
     {D_N}^{\e_1' \dots \e_m'}_{\e_1 \dots \e_m}
           (\x_1, \dots, \x_m|\k, \a) =
	\frac{\<\k + \a| T^{\e_1'}_{\e_1} (\x_1, \k) \dots
	                 T^{\e_m'}_{\e_m} (\x_m, \k) |\k\>}
             {\<\k + \a|\prod_{j=1}^m t (\x_j,\k)|\k\>} \epc
\end{equation}
which is in fact an inhomogeneous and `$\a$-twisted' version of the
usual density matrix.

In the finite length case (\ref{tdistr}) with twist angle $\PH$ the
expectation value in the ground state $|\PH\>$ of any operator
$X_{[1,m]}$ acting non-trivially only on the first $m$ lattice
sites is \cite{DGHK07}
\begin{equation}
     \frac{\<\PH|X_{[1,m]}|\PH\>}{\<\PH|\PH\>} =
        \lim_{\a \rightarrow 0}\, \lim_{\n_j \rightarrow \h/2}
	\tr_{1, \dots, m} \bigl\{
	    {D_N} (\x_1, \dots, \x_m|- \PH/\g, \a)\, X_{[1,m]}
	    \bigr\} \epp
\end{equation}
In the finite temperature case (\ref{qtmdistr}) we use that the
right hand side of (\ref{defdens}) stays form invariant under
the transformation (\ref{ttqtm}) which replaces all objects
relating to the ordinary transfer matrix with the corresponding
objects relating to the quantum transfer matrix. Hence, from
\cite{GKS05},
\begin{multline}
     \<X_{[1,m]}\>_{T, h} =
        \lim_{L \rightarrow \infty}
	\frac{\tr_{1, \dots, L} \bigl\{ \re^{- \be H_L (0) + h S_{[1,L]}/T}%
	      X_{[1,m]} \bigr\}}{Z_L} \\[1ex] =
        \lim_{\a \rightarrow 0}\, \lim_{\n_j \rightarrow 0}
        \lim_{N \rightarrow \infty}
	\tr_{1, \dots, m} \bigl\{
	    {D_N} (\x_1, \dots, \x_m|h/(2\h T), \a)\, X_{[1,m]}
	    \bigr\} \epp
\end{multline}

The density matrix (\ref{defdens}) allows for reduction from the
left and from the right expressed by
\begin{subequations}
\label{redu}
\begin{align}
     \tr_1 \bigl\{ D_N (\x_1, \dots, \x_m|\k, \a) q^{\a \s_1^z} \bigr\} & =
        \r (\x_1) D_N (\x_2, \dots, \x_m|\k, \a) \epc \\[1ex]
     \tr_m \bigl\{ D_N (\x_1, \dots, \x_m|\k, \a) \bigr\} & =
        D_N (\x_1, \dots, \x_{m-1}|\k, \a) \epc
\label{reductionD}
\end{align}
\end{subequations}
where
\begin{equation} \label{defrho}
     \r (\z) = \frac{\La (\z, \k + \a)}{\La (\z, \k)} \epp
\end{equation}
The function $\r$ plays an important role in \cite{JMS08pp}. As we
shall see below it is also important for the formulation of a
multiple integral formula for the density matrix and is the only
non-trivial one-point function for finite $\a$. In the temperature case
with $\k = h/(2\h T)$ we have
\enlargethispage{3ex}
\begin{equation}
     \r(1) = 1 + m(T, h)2 \h \a + {\cal O} (\a^2) \epc
\end{equation}
where $m(T, h)$ is the magnetization.

\begin{figure}
    \centering
    \includegraphics{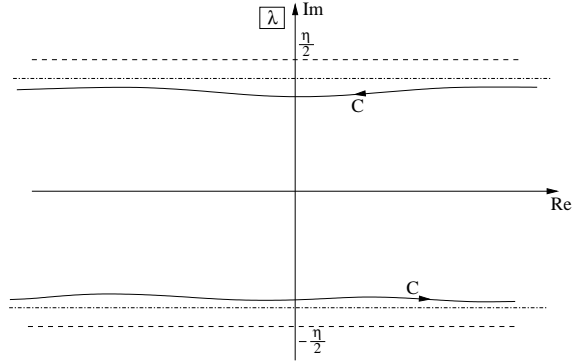}
    \caption{\label{fig:cont} The canonical contour ${\cal C}$ surrounds
             the real axis in counterclockwise manner inside the
             strip $- \frac{\g}{2} < \Im \la < \frac{\g}{2}$.} 
\end{figure}

In the temperature case as well as in the finite length case and
in certain inhomogeneous generalizations of both cases the function
$\r$ can be expressed in terms of an integral over certain auxiliary
functions (see e.g.\ \cite{GKS04a,DGHK07}),
\begin{equation} \label{rhoint}
     \r(\z) = q^\a \exp \biggl\{
                  \int_{\cal C} \frac{\rd \m}{2 \p \i} \: \re (\m - \la)
		  \ln \biggl[ \frac{1 + \fa (\m, \k + \a)}
		                   {1 + \fa (\m, \k )} \biggr] \biggr\}
				   \epp
\end{equation}
Here $\re(\la)$ is the `bare energy'
\begin{equation}
     \re(\la) = \cth(\la) - \cth(\la + \h)
\end{equation}
and $\fa(\la, \k)$ is the solution of a non-linear integral equation
with integration kernel
\begin{equation} \label{kernel}
     K(\la) = \cth(\la - \h) - \cth(\la + \h) \epp
\end{equation}
In the finite length case this equation reads
\begin{multline} \label{nliefin}
     \ln (\fa (\la, \k)) = \\ (N - 2\k) \h + \sum_{j=1}^N
		           \ln \biggl[ \frac{\sh (\la - \be_j)}
		                   {\sh (\la - \be_j + \h)} \biggr]
                           - \int_{\cal C} \frac{\rd \m}{2 \p \i}
			     K(\la - \m) \ln (1 + \fa (\m, \k )) \epp
\end{multline}
Equations (\ref{rhoint}) and (\ref{nliefin}) are still valid if the
$\be_j$ are not precisely those of equation (\ref{tdistr}), but are close
to $\h/2$ with $\Im \be_j = \g/2$. The contour of integration to be used
in (\ref{rhoint}) and (\ref{nliefin}) is shown in figure \ref{fig:cont}.
In the temperature case the non-linear integral equation has a similar
structure, but the driving term is different. Suppose that for $j = 1,
\dots, N/2$ the $\be_{2j-1}$ are close to $\h$, whereas the $\be_{2j}$
are close to $0$. Then
\begin{multline} \label{nlietem}
     \ln (\fa (\la, \k)) = - 2 \k \h \\ + \sum_{j=1}^{N/2}
	  \ln \biggl[ \frac{\sh(\la - \be_{2j})
	                    \sh(\la - \be_{2j-1} + 2\h)}
			   {\sh (\la - \be_{2j} + \h)
			    \sh(\la - \be_{2j-1} + \h)} \biggr]
                           - \int_{\cal C} \frac{\rd \m}{2 \p \i}
			     K(\la - \m) \ln (1 + \fa (\m, \k )) \epp
\end{multline}

We presented both equations (\ref{nliefin}) and (\ref{nlietem}) in
inhomogeneous form, since we shall need this later, when comparing
with \cite{JMS08pp}. Note, however, that the homogeneous limit is
trivial in both cases and that, moreover, the Trotter limit
can be performed in (\ref{nlietem}). Then
\begin{equation} \label{nliefinhom}
     \ln (\fa (\la, \k)) = (N - 2\k) \h
        + N \ln \biggl[ \frac{\sh (\la - \h/2)}
	                     {\sh (\la + \h/2)} \biggr]
        - \int_{\cal C} \frac{\rd \m}{2 \p \i}
		        K(\la - \m) \ln (1 + \fa (\m, \k )) \epp
\end{equation}
in the finite length case and
\begin{equation} \label{nlietemhom}
     \ln (\fa (\la, \k)) = - 2\k \h - \frac{2J \sh(\h) \re (\la)}{T}
        - \int_{\cal C} \frac{\rd \m}{2 \p \i}
		        K(\la - \m) \ln (1 + \fa (\m, \k ))
\end{equation}
in the temperature case and in the Trotter limit. Equations
(\ref{nliefinhom}) and (\ref{nlietemhom}) are what we call the
$\fa$-form of the non-linear integral equation. There is another
so-called $\fb \fbq$-form \cite{Kluemper92,DGHK07} which is more
convenient for an accurate calculation of the numerical values
of the functions.

\section{The multiple integral representation of the density matrix}
\label{sec:multint}
In appendix \ref{app:dermult} we derive the following multiple integral
representation for the elements of the density matrix.
\begin{multline} \label{multint}
     {D_N}^{\e_1' \dots \e_m'}_{\e_1 \dots \e_m}
           (\x_1, \dots, \x_m|\k, \a) =
	   \biggl[ \prod_{j=1}^p \int_{\cal C} \rd m(\la_j) \:
	          F^+_{\ell_j} (\la_j) \biggr]
	   \biggl[ \prod_{j=p+1}^m \int_{\cal C} \rd \overline{m}(\la_j) \:
	          F^-_{\ell_j} (\la_j) \biggr] \\[1ex]
           \frac{\det_{j, k = 1, \dots, m} \bigl[- G(\la_j, \n_k) \bigr]}
	        {\prod_{1 \le j < k \le m} \sh(\la_j - \la_k - \h)
		 \sh(\n_k - \n_j)} \epc
\end{multline}
where we have used the notation
\begin{align}
     \rd m(\la) &
        = \frac{\rd \la}{2 \p \i \, \r(\z) (1 + \fa (\la, \k))} \epc \qd
     \rd \overline{m} (\la)
        = \fa (\la, \k) \rd m(\la) \epc \\[2ex] \notag
     F_{\ell_j}^\pm (\la) &
        = \prod_{k=1}^{\ell_j - 1} \sh(\la - \n_k)
	  \prod_{k=\ell_j + 1}^m \sh(\la - \n_k \mp \h) \epc \qd
     \ell_j = \begin{cases}
                 \e_j^+ & j = 1, \dots, p \\
		 \e_{m - j + 1}^- & j = p + 1, \dots, m
               \end{cases}
\end{align}
with $\e_j^+$ the $j$th plus in the sequence $(\e_j)_{j=1}^m$, $\e_j^-$
the $j$th minus sign in the sequence $(\e_j')_{j=1}^m$ and $p$ the
number of plus signs in $(\e_j)_{j=1}^m$. The function $G$ is new here.
It is defined as the solution of the linear integral equation
\begin{equation} \label{newg}
     G(\la, \n) = q^{-\a} \cth(\la - \n - \h) - \r (\x) \cth (\la - \n) 
                  + \int_{\cal C} \rd m(\m) K_\a (\la - \m) G(\m, \n) \epc
\end{equation}
where $\x = \re^\n$, and the kernel
\begin{equation}
     K_\a (\la) = q^{- \a} \cth (\la - \h) - q^\a \cth (\la + \h)
\end{equation}
is a deformed version of (\ref{kernel}).

Equation (\ref{multint}) is a generalization to finite $\a$ of the multiple
integral formulae first derived in \cite{GKS05,DGHK07}. To simplify
the notation we shall sometimes suppress the dependence of the density
matrix elements on $\k$ and $\a$.
\section{The case m = 1}
For $m = 1$ there are only two non-vanishing density matrix elements.
They are related to the function $\r$ by the reduction relations
(\ref{redu}) which imply that
\begin{equation}
     \begin{pmatrix} D^+_+ (\x) \\ D^-_- (\x) \end{pmatrix}
        = \frac{1}{q^\a - q^{-\a}}
	  \begin{pmatrix}
	     - q^{-\a} & \mspace{14.mu} 1 \\ q^\a & - 1
	  \end{pmatrix}
	  \begin{pmatrix} 1 \\ \r(\x) \end{pmatrix} \epp
\end{equation}
When we insert equation (\ref{multint}) for $m = 1$ here, we do not
obtain an independent equation, but rather an interesting identity
for $\r$ (recall that $\r$ appears in the measure),
\begin{equation} \label{grhoid}
     \r (\x) = q^{-\a} - (q^\a - q^{-\a}) \int_{\cal C} \rd m(\m) G(\m, \n)
               \epp
\end{equation}
It allows us to calculate the asymptotic behaviour of the
function $G$,
\begin{equation}
     \lim_{\Re \la \rightarrow \pm \infty} G(\la, \n) = 0 \epp
\end{equation}
\section{Factorization of the density matrix for m = 2}
The factorization of the multiple integrals for the ground state
density matrix was discovered in \cite{BoKo01}. In that case the
integrand consists of explicit functions whose analytic properties
were used in the calculation. In the finite temperature case a
different factorization technique had to be invented. As was demonstrated 
in \cite{BGKS06} the linear integral equation for the function G,
appropriately used under the multiple integral, can be viewed as the
source of the factorization, at least for the special case of the
isotropic chain at $\a = 0$. For the XXZ chain outside the isotropic
point and without the disorder parameter $\a$, however, that trick
does not work anymore. Here we shall see that a finite $\a$ allows us
to perform the factorization of the density matrix in much the same
way as in \cite{BGKS06}.

Let us consider $m = 2$ in (\ref{multint}). There are six non-vanishing
matrix elements in this case, one for $p = 0$, four for $p = 1$ and one
for $p = 2$. We shall concentrate on the case $p = 1$, since the matrix
elements for $p = 0$ or $ 2$ can be obtained from those for $p = 1$ by
means of the reduction relation (\ref{redu}). After substituting
$w_j = \re^{2 \m_j}$ and $\x_j = \re^{\n_j}$, $j = 1, 2$, the corresponding
integrals are all of the form
\begin{equation} \label{intm2}
     {\cal I} = \frac{1}{\x_2^2 - \x_1^2}
        \int_{\cal C} \rd m(\m_1) \int_{\cal C} \rd \overline{m} (\m_2)
	\det \bigl[ G(\m_j, \n_k) \bigr] r(w_1, w_2) \epc
\end{equation}
where
\begin{equation}
     r(w_1, w_2) = \frac{p(w_1, w_2)}{w_1 - q^2 w_2} \epc \qd
     p(w_1, w_2) = c_0 w_1 w_2 + c_1 w_1 + c_2 w_2 + c_3 \epp
\end{equation}
The coefficients $c_j$ are different for the four different matrix
elements. They are listed in table \ref{tab:pcoeff}.
\renewcommand{\arraystretch}{1.5}
\begin{table}[t]
\begin{minipage}{\linewidth}
    \centering
    \begin{tabular}{ccccc}
      \toprule
	$\begin{smallmatrix} \e_1' & \e_2' \\ \e_1 & \e_2
	 \end{smallmatrix}$
        & $c_0$ & $c_1$  & $c_2$ & $c_3$ \\
      \midrule
	$\begin{smallmatrix} + & - \\ + & - \end{smallmatrix}$
	& 1 & $- \x_1^2$ & $- q^2 \x_2^2$ & $q^2 \x_1^2 \x_2^2$ \\
	$\begin{smallmatrix} - & + \\ - & + \end{smallmatrix}$
	& $q^2$  & $- \x_2^2$ & $- q^2 \x_1^2$ & $\x_1^2 \x_2^2$ \\
	$\begin{smallmatrix} + & - \\ - & + \end{smallmatrix}$
	& $q \x_2/\x_1$ & $- q \x_1 \x_2$ & $- q \x_1 \x_2$
	& $q \x_1^3 \x_2$ \\
	$\begin{smallmatrix} - & + \\ + & - \end{smallmatrix}$
	& $q \x_1/\x_2$ & $- q^{-1} \x_1 \x_2$ & $- q^3 \x_1 \x_2$
	& $q \x_1 \x_2^3$ \\
      \bottomrule
    \end{tabular}
    \caption{\label{tab:pcoeff} The coefficients of the polynomial $p$.}
\end{minipage}
\end{table}
\renewcommand{\arraystretch}{1.1}

Inserting
\begin{equation}
     \rd \overline{m} (\m) = \frac{\rd \la}{2 \p \i \r(\re^\m)}
                             - \rd m(\m)
\end{equation}
into (\ref{intm2}) and taking into account that $\r (\re^\m)$ is
analytic and non-zero inside $\cal C$ we obtain
\begin{multline} \label{ij}
     {\cal I} (\x_2^2 - \x_1^2) = - \int_{\cal C} \rd m(\m) \,
        \det \begin{pmatrix} G(\m, \n_1) & G(\m, \n_2) \\
	                     r(w, \x_1^2) & r(w, \x_2^2) \end{pmatrix}
			     \\[1ex]
        - \int_{\cal C} \rd m(\m_1) \int_{\cal C} \rd m (\m_2)
	  \det \bigl[ G(\m_j, \n_k) \bigr] r(w_1, w_2) \epc
\end{multline}
where $w = \re^{2 \m}$. Here the first integral is already factorized.
Under the second integral the integration measures now appear
symmetrically. Hence, we may replace $r(w_1, w_2)$ by $(r(w_1, w_2) -
r(w_2, w_1))/2$.

Following \cite{BGKS06} we want to use the integral equation (\ref{newg})
under the second integral in (\ref{ij}). This is possible if rational
functions $F(w_1, w_2)$  and $g(w)$ exist, such that
\begin{equation} \label{decr}
     r(w_1, w_2) - r(w_2, w_1) = F(w_1, w_2) +
        g(w_1) K_\a (\m_1 - \m_2) - g(w_2) K_\a (\m_2 - \m_1) \epc
\end{equation}
and the antisymmetric function $F(w_1, w_2)$ is a sum of factorized
functions in $w_1$ and $w_2$. Then $F$ considered as a function of
$w_1$ cannot have poles whose position depends on $w_2$. In particular,
the residue at $w_1 = q^2 w_2$ must vanish. Using this in (\ref{decr})
with the explicit forms of $r$ and $K_\a$ inserted we obtain a
difference equation for $g$,
\begin{equation} \label{diffg}
     g(q^2 w) y^{-1} - g(w) y = \frac{p(q^2 w, w)}{2 q^2 w} \epp
\end{equation}
Here $y = q^\a$. Clearly this equation has a solution of the form
\begin{equation}
     g(w) = g_+ w + g_0 + \frac{g_-}{w} \epp
\end{equation}
The coefficients are easily obtained by substituting the latter expression
into (\ref{diffg}),
\begin{equation}
     g_+ = \frac{c_0 y}{2(q^2 - y^2)} \epc \qd
     g_- = \frac{c_3 y}{2(1 - q^2 y^2)} \epc \qd
     g_0 = \frac{(c_1 + q^{-2} c_2) y}{2(1 - y^2)} \epp
\end{equation}
Substituting $g$ back into (\ref{decr}) we obtain $F(w_1, w_2) = f(w_1)
- f(w_2)$, where
\begin{equation}
     f(w) = (y - y^{-1})\Bigl(g_+ w - \frac{g_-}{w}\Bigr) \epp
\end{equation}
Consequentially
\begin{equation}
     r(w_1, w_2) = f(w_1) + g(w_1) K_\a (\m_1 - \m_2)
                   + \text{symmetric function.}
\end{equation}

With this we can factorize the second integral in (\ref{ij}) by means
of the integral equation (\ref{newg}),
\begin{multline} \label{ifirstfac}
     \int_{\cal C} \rd m(\m_1) \int_{\cal C} \rd m (\m_2)
        \det \bigl[ G(\m_j, \n_k) \bigr] r(w_1, w_2) \\
        = (y - y^{-1})
	  \det \begin{pmatrix}
	     g_+ \ph_+ (\n_1) - g_- \ph_- (\n_1) &
	     g_+ \ph_+ (\n_2) - g_- \ph_- (\n_2) \\
	     \ph_0 (\n_1) & \ph_0 (\n_2)
	  \end{pmatrix} \\[1ex] +
          \int_{\cal C} \rd m(\m) \,
	  \det \begin{pmatrix}
	       G(\m, \n_1) & G(\m, \n_2) \\
	       g(w) H(\m, \n_1; y^{-1}) & g(w) H(\m, \n_2; y^{-1})
	  \end{pmatrix} \epc
\end{multline}
where
\begin{subequations}
\begin{align}
     \ph_j (\n) & = \int_{\cal C} \rd m(\m) \, w^j G(\m, \n) \epc \qd
                    j = +, 0, - \epc \\
     H(\m, \n; y^{-1}) &
        = \r(\x) \cth (\m - \n) - y^{-1} \cth(\m - \n - \h) \epp
\end{align}
\end{subequations}
Finally we substitute (\ref{ifirstfac}) into (\ref{ij}) and further
simplify the resulting expression using the identities
\begin{subequations}
\begin{align}
     g(w) H(\m, \n; y^{-1}) & = g(\x^2) H(\m, \n; y)
        - \frac{p(q^2 \x^2, \x^2)}{2 q^2 \x^2} \cth(\m - \n - \h)
	\notag \\
	& \mspace{36.mu} - \ph_0 (\n)  \bigl(f(w) - f(\x^2)\bigr)
	  + \frac{y^{-1}}{y - y^{-1}} \bigl(f(\x^2) - f(q^2 \x^2)\bigr)
	  \epc \\[1ex]
     r(w, \x^2) & = \frac{p(q^2 \x^2, \x^2)}{2 q^2 \x^2} \cth(\m - \n - \h)
		  - \frac{p(- q^2 \x^2, \x^2)}{2 q^2 \x^2} \epp
\end{align}
\end{subequations}
Then
\begin{multline} \label{innerfac}
     {\cal I} =
         \frac{g(\x_2^2) \Ps (\x_2, \x_1) - g(\x_1^2) \Ps (\x_1, \x_2)}
              {\x_2^2 - \x_1^2}
         + \frac{(c_1 - q^{-2} c_2)(\r(\x_1) - \r(\x_2))}
	        {2(\x_2^2 - \x_1^2)(y - y^{-1})} \\[1ex]
         + \frac{(y^{-1} - \r(\x_1))(y - \r(\x_2)) f(\x_2^2)
	         - (y^{-1} - \r(\x_2))(y - \r(\x_1)) f(\x_1^2)}
	        {(\x_2^2 - \x_1^2)(y - y^{-1})^2} \epc
\end{multline}
where
\begin{equation} \label{Psi}
     \Ps (\x_1, \x_2) =
        \int_{\cal C} \rd m(\m) G(\m, \n_2)
	\bigl(q^\a \cth(\m - \n_1 - \h) - \r(\x_1) \cth(\m - \n_1) \bigr)
	\epp
\end{equation}

Equation (\ref{innerfac}) determines the four density matrix elements
for $p = 1$ in factorized form. Note that the matrix elements depend
on only two transcendental functions $\r$ and $\Ps$. The remaining
two non-zero density matrix elements for $m = 2$ follow from
(\ref{innerfac}) by means of the reduction relations (\ref{redu}),
\begin{subequations}
\begin{align}
     D^{++}_{++} (\x_1, \x_2) &
        = \frac{\r(\x_1) - y^{-1}}{y - y^{-1}}
	  - D^{+-}_{+-} (\x_1, \x_2) \epc \\
     D^{--}_{--} (\x_1, \x_2) &
        = \frac{y - \r(\x_1)}{y - y^{-1}} - D^{-+}_{-+} (\x_1, \x_2) \epp
\end{align}
\end{subequations}
We shall give a fully explicit matrix representation of the factorized
density matrix for $m = 2$ below, after we have introduced the function
$\om$.

\section{The function $\om$} \label{sec:om}
In the recent work \cite{JMS08pp} is was shown that the correlation
functions defined by the inhomogeneous and $\a$-twisted density
matrix (\ref{defdens}) factorize and can all be expressed in terms
of only two transcendental functions, the function $\r$ entering the
reduction relations (\ref{redu}) and another function $\om$ which
in \cite{JMS08pp} was defined as the expectation value of a product
of two creation operators and was represented by a determinant formula.
The approach of \cite{JMS08pp} is slightly different from ours here
in that the lattice used in \cite{JMS08pp} is homogeneous in `horizontal
direction' (all the $\x$s in (\ref{defdens}) are taken to be 1 from
the outset). For the ground state both cases lead to the same function
$\om$ (see section 5.3 and 5.4 of \cite{BJMST08app}). In particular,
in the inhomogeneous case, following sections 5.1 and 5.3 of
\cite{BJMST08app}, we have\footnote{More precisely this function
was denoted $(\om_0 - \om)(\x_1/\x_2, \a)$ in \cite{BJMST08app}.}
\begin{equation} \label{defom}
     \om (\x_1, \x_2)
        = - \bigl\langle
	     \cv^*_{[1,2]} (\x_2,\a) \bv^*_{[1,2]}(\x_1,\a - 1) (1)
	    \bigr\rangle \epp
\end{equation}

Replacing the vacuum expectation value by the expectation value
calculated with the density matrix (\ref{defdens}) we take (\ref{defom})
as our definition of the function $\om$. In our case $\om$ depends on
two twist parameters $\k$ and $\a$. We indicate this by writing 
$\om (\x_1, \x_2| \k, \a)$. The construction of the operators
$\bv^*_{[1,2]}$ and $\cv^*_{[1,2]}$ is explained in \cite{BJMST08app}.
For the product needed in (\ref{defom}) we find the explicit expression
\begin{align} \label{c2b11}
     \x^{-\a} \cv^*_{[1,2]} & (\x_2, \a) \bv^*_{[1,2]} (\x_1,\a - 1) (1) =
        \notag \\[1ex] &
        \biggl( \frac{q^{\a - 1} \x^{-1}}{q \x - q^{-1} \x^{-1}} -
                \frac{q^{1 - \a} \x^{-1}}{q^{-1} \x - q \x^{-1}} +
                \frac{q^\a - q^{- \a}}{2} \biggr) \s^z \otimes \s^z
        \notag \\[1ex] & +
        \frac{q^\a - q^{- \a}}{2}
        \biggl( \frac{q^{-1} \x^{-1}}{q \x - q^{-1} \x^{-1}} -
                \frac{q \x^{-1}}{q^{-1} \x - q \x^{-1}} \biggr)
                \bigl( I_2 \otimes \s^z - \s^z \otimes I_2 \bigr)
        \notag \\[1ex] & +
        2 \biggl( \frac{q^\a}{q \x - q^{-1} \x^{-1}} -
                  \frac{q^{- \a}}{q^{-1} \x - q \x^{-1}} \biggr)
                  \bigl( \s^+ \otimes \s^- + \s^- \otimes \s^+ \bigr)
        \notag \\[1ex] & +
        (q^\a - q^{- \a})
        \biggl( \frac{1}{q \x - q^{-1} \x^{-1}} +
                \frac{1}{q^{-1} \x - q \x^{-1}} \biggr)
                \bigl( \s^+ \otimes \s^- - \s^- \otimes \s^+ \bigr) \epc
\end{align}
where $\x = \x_1/\x_2$. Inserting this into (\ref{defom}) and calculating
the average with the factorized two-site density matrix of the previous
section we obtain
\begin{equation} \label{ompsi}
     \om(\x_1, \x_2|\k, \a) = 2 \x^\a \Ps(\x_1, \x_2) - \D \ps(\x)
                       + 2 \bigl( \r(\x_1) - \r(\x_2) \bigr) \ps(\x) \epp
\end{equation}
Here we adopted the notation from \cite{BJMST08app},
\begin{equation} \label{defpsi}
     \ps(\x) = \frac{\x^\a (\x^2 + 1)}{2(\x^2 - 1)} \epc
\end{equation}
and $\D$ is the difference operator whose action on a function $f$
is defined by $\D f(\x) = f(q \x) - f(q^{-1} \x)$.

The remaining part of this section is devoted to the exploration of
the properties of~$\om$. First of all we substitute $\om$ back into
the equation for the two-site density matrix, which can then be
expressed entirely in terms of $\om$ and a function
\begin{equation}
     \ph (\z|\k, \a) = \frac{\ch (\a \h) - \r(\z)}{\sh(\a \h)}
\end{equation}
which is sometimes more convenient than the function $\r$ itself.
We obtain
\begin{align}
     D_N & (\x_1, \x_2|\k, \a) = \4 I_2 \otimes I_2 \notag \\
        &  - \frac{1}{4(q^{\a - 1} - q^{1 - \a})}
          \biggl( \frac{\x^{1 - \a} \om_{12} - \x^{\a - 1} \om_{21}}
                       {\x - \x^{-1}} +
                  \frac{\ph_1 \ph_2 (q^\a - q^{- \a})}{2} \biggr)
          \notag \\ & \qd
          \biggl( \frac{q - q^{-1}}{2} I_2 \otimes \s^z
                - \frac{q + q^{-1}}{2} \s^z \otimes \s^z
                + \x^{-1} \, \s^+ \otimes \s^- + \x \, \s^- \otimes \s^+
          \biggr) \notag \\[1ex]
        &  - \frac{1}{4(q^{\a + 1} - q^{- \a - 1})}
          \biggl( \frac{\x^{- \a - 1} \om_{12} - \x^{\a + 1} \om_{21}}
                       {\x - \x^{-1}} +
                  \frac{\ph_1 \ph_2 (q^\a - q^{- \a})}{2} \biggr)
          \notag \\ & \qd
          \biggl( - \frac{q - q^{-1}}{2} I_2 \otimes \s^z
                - \frac{q + q^{-1}}{2} \s^z \otimes \s^z
                + \x \, \s^+ \otimes \s^- + \x^{-1} \, \s^- \otimes \s^+
          \biggr) \notag \\[1ex]
        &  - \frac{\x^{-\a} \om_{12} - \x^\a \om_{21}}
                  {4(\x - \x^{-1})(q^\a - q^{- \a})}
          \bigl( (\x + \x^{-1}) \s^z \otimes \s^z
                - (q + q^{-1})(\s^+ \otimes \s^- +  \s^- \otimes \s^+)
          \bigr) \notag \\[1ex]
        & - \4 \bigl(\ph_1 \, \s^z \otimes I_2
                     + \ph_2 \, I_2 \otimes \s^z \bigr)
          - \frac{q - q^{-1}}{4(\x - \x^{-1})} (\ph_1 - \ph_2)
            (\s^+ \otimes \s^- - \s^- \otimes \s^+) \epc
\end{align}
where we introduced the abbreviations $\om_{jk} = \om(\x_j, \x_k| \k, \a)$
and $\ph_j = \ph(\x_j| \k, \a)$.

For the limit $\a \rightarrow 0$ the properties of the functions $\ph$
and $\om$ with respect to negating $\k$ and $\a$ are important. They
follow from the fact that the $R$-matrix is invariant under spin reversal,
\begin{equation} \label{rspinrev}
     R(\la) = (\s^x \otimes \s^x) R(\la) (\s^x \otimes \s^x) \epp
\end{equation}
Introducing the spin reversal operator $J = \s_1^x \dots \s_N^x$
we conclude with (\ref{rspinrev}) that
\begin{equation}
     T_a (\z, - \k) = \s_a^x J \, T_a (\z, \k) \, J \s_a^x \epp
\end{equation}
It follows that $t(\z, -\k) = J t(\z, \k) J$. Hence,
\begin{subequations}
\begin{align}
     J |\k\> & = |-\k\> \epc \\[1ex] \La (\z, \k) & = \La (\z, -\k) \epp
        \label{evinv}
\end{align}
\end{subequations}
The latter two equations used in the definition (\ref{defdens}) of
the $\a$-twisted density matrix imply that
\begin{equation} \label{densrevers}
     D_N (\x_1, \dots, \x_m|- \k, - \a) = (\s^x)^{\otimes m} \,
        D_N (\x_1, \dots, \x_m|\k, \a) \, (\s^x)^{\otimes m} \epp
\end{equation}

From (\ref{defrho}), (\ref{evinv}) we obtain the relation
\begin{equation}
     \ph(\z|- \k, - \a) = - \ph(\z| \k, \a) \epp
\end{equation}
Equation (\ref{densrevers}) together with (\ref{defom})-(\ref{ompsi})
and the expressions for the density matrix elements of the previous
section implies that
\begin{equation}
     \om(\x_1, \x_2|\k, \a) = \om(\x_2, \x_1|- \k, - \a) \epp
\end{equation}

Our next step is to verify that the function $\om$ given by the formula
(\ref{ompsi}) satisfies a property called the 'normalization condition'
by the authors of \cite{JMS08pp} (see equation (6.10) there). So we come
back to the case of finite Trotter number $N$ with arbitrary inhomogeneity
parameters $\be_j$, $j = 1, \dots, N$ as it is written in (\ref{defmono}).
We shall also use multiplicative parameters $\tau_j=e^{\be_j}$.

We consider the normalization condition in the following form 
\begin{multline} \label{norm}
     \bigl(\om(\z,\xi|\k, \a)
        + {\overline{D}}_{\z}{\overline D}_{\xi}\Delta_{\z}^{-1}
          \psi(\z/\xi)\bigr)\bigr|_{\z=\tau_j} + \\
        +\rho(\tau_j) \bigl(\om(\z,\xi|\k, \a)
        +{\overline D}_{\z}{\overline D}_{\xi}\Delta_{\z}^{-1}
        \psi(\z/\xi)\bigr)\bigr|_{\z=q^{-1}\tau_j} = 0 \epc
\end{multline}
$j = 1, \dots N$, which can be obtained from the integral in (6.10) of
\cite{JMS08pp} by taking the residues and using the TQ-relation (4.2)
of that paper. Also let us recall the definition 
\begin{equation}
     {\overline{D}}_{\z} g(\z) = g(q\z)+g(q^{-1}\z)-2\rho(\z)g(\z) \epp
\label{D}
\end{equation}
Actually, (6.10) of \cite{JMS08pp} comprises one more equation related
to the residue at $\z^2 = 0$. This case needs separate treatment and will
be discussed below.

First we use the following difference equation for the function 
$\Psi$ defined by (\ref{Psi}),
\begin{align} \label{eqPsi}
     \Psi&(\xi_1,\xi_2)+\rho(\xi_1)q^{-\a}\Psi(q^{-1}\xi_1, \x_2)=
        \frac{G(\nu_1,\nu_2)}{1+\bar\fa(\nu_1,\kappa)}
       -\rho(\xi_1)q^{-\a}\frac{G(\nu_1-\eta,\nu_2)}
                                {1+\fa(\nu_1-\eta,\kappa)} \notag \\[1ex]
     &+\rho(\xi_2)\cth(\nu_1-\nu_2)-q^{-\a}\cth(\nu_1-\nu_2-\eta)
      \notag \\[1ex]
     &- q^{- \a} \bigl(\rho(\xi_1)\rho(q^{-1}\xi_1)-1\bigr)
       \int_{\cal C} \rd m(\mu) G(\mu,\nu_2)\cth(\mu-\nu_1+\eta) \epc
\end{align}
where $\faq = 1/\fa$ by definition.
This equation is the result of an analytical continuation defined for
$\Psi(q^{-1} \x_1, \x_2)$ through an appropriate deformation of the
integration contour in (\ref{Psi}). Some simplifications occur in the
limit $\nu_1 \rightarrow \be_j$ or equivalently $\xi_1 \rightarrow \tau_j$,
namely, since $\fa(\be_j,\kappa)=\bar\fa(\be_j-\eta,\kappa)=0$ or
$\bar\fa(\be_j,\kappa)=\fa(\be_j-\eta,\kappa)=\infty$, the first two terms
in the right hand side of (\ref{eqPsi}) do not contribute. Then we have
\[
     \rho(\tau_j)\rho(q^{-1}\tau_j) = 
       \frac{Q^-(q^{-1}\tau_j;\kappa+\a)Q^+(\tau_j;\kappa)}
            {Q^-(\tau_j;\kappa+\a)Q^+(q^{-1}\tau_j;\kappa)}\cdot
       \frac{Q^-(\tau_j;\kappa+\a)Q^+(q^{-1}\tau_j;\kappa)}
            {Q^-(q^{-1}\tau_j;\kappa+\a)Q^+(\tau_j;\kappa)} = 1
\]
with the $Q$-functions $Q^\pm$ defined in \cite{JMS08pp}. This means
that also the last term in the right hand side of (\ref{eqPsi}) does
not contribute. Hence, we obtain
\begin{equation} \label{eqPsi1}
     \Psi(\tau_j, \x_2) + \rho(\tau_j)q^{-\a}\Psi(q^{-1}\tau_j, \x_2) =
        \rho(\xi_2) \cth(\be_j-\nu_2)-q^{-\a} \cth(\be_j-\nu_2-\eta) \epp
\end{equation}
Note that the right hand side is, up to the sign, equal to the driving
term in the integral equation (\ref{newg}) for $G$.

If we take the formula (\ref{ompsi}) and use (\ref{eqPsi1}) then, after
some algebra, we obtain 
\begin{align} \label{eqom}
     \om(&\tau_j,\xi_2|\kappa,\a)
        +\rho(\tau_j)\om(q^{-1}\tau_j,\xi_2|\kappa,\a) = \notag \\[1ex]
     &-{\bigl(\Delta_{\z}\psi(\z/\xi_2)\bigr)}\bigr|_{\z=\tau_j}-
        \rho(\tau_j){\bigl(\Delta_{\z}\psi(\z/\xi_2)\bigr)}
        \bigr|_{\z=q^{-1}\tau_j} \notag \\[1ex]
     &+2(\rho(\tau_j)+\rho(\xi_2))\;\psi(\tau_j/\xi_2)
      -2(1+\rho(\tau_j)\rho(\xi_2))\;\psi(q^{-1}\tau_j/\xi_2) \epp
\end{align}
Now we need to check that this equation is equivalent to (\ref{norm}).
To this end we should verify the following equality 
\begin{multline} \label{equal}
     \bigl( {\overline{D}}_{\z} {\overline D}_{\xi}
             \Delta_{\z}^{-1}\psi(\z/\xi)\bigr)\bigr|_{\z=\tau_j}
     +\rho(\tau_j) \bigl({\overline D}_{\z}{\overline D}_{\xi}
      \Delta_{\z}^{-1}\psi(\z/\xi) \bigr) \bigr|_{\z=q^{-1}\tau_j}
     = \\[1ex]
     \bigl( \Delta_{\z}\psi(\z/\xi_2) \bigr) \bigr|_{\z=\tau_j} +
     \rho(\tau_j) \bigl( \Delta_{\z} \psi(\z/\xi_2) \bigr)
     \bigr|_{\z=q^{-1}\tau_j} \\
     -2(\rho(\tau_j)+\rho(\xi_2)) \psi(\tau_j/\xi_2)
     +2(1+\rho(\tau_j)\rho(\xi_2))\psi(q^{-1}\tau_j/\xi_2) \epp
\end{multline}
Using the definition (\ref{D}) we come after a little algebra to the
following expression for an arbitrary function $g(\z)$ 
\begin{multline} \label{actdbar}
     {\overline{D}}_{\z}{\overline D}_{\xi}\;g(\z/\xi)=
        \Delta_{\z}^2\;g(\z/\xi)
     + 4(1-\rho(\z))\;(1-\rho(\xi))\;g(\z/\xi) \\
     -2(\rho(\z)+\rho(\xi))\;(g(q\z/\xi)+g(q^{-1}\z/\xi)-2g(\z/\xi)) \epp
\end{multline}
Now take 
\begin{multline}
     \bigl({\overline{D}}_{\z} {\overline D}_{\xi}
      \;g(\z/\xi)\bigr)\bigr|_{\z=\tau_j}+
      \rho(\tau_j) \bigl( {\overline D}_{\z} {\overline D}_{\xi}
      \;g(\z/\xi)\bigr) \bigr|_{\z=q^{-1}\tau_j} = \\[1ex]
     \bigl(\Delta_{\z}^2\;g(\z/\xi)\bigr)\bigr|_{\z=\tau_j}
      +\bigl(\Delta_{\z}^2 \;g(\z/\xi)\bigr)\bigr|_{\z=q^{-1}\tau_j} -
      2(\rho(\tau_j)+\rho(\xi)) \bigl(\Delta_{\z}\;g(\z/\xi)\bigr)
      \bigr|_{\z=\tau_j} \\[1ex] + 2(1+\rho(\tau_j)\rho(\xi))
      \bigl(\Delta_{\z}\;g(\z/\xi)\bigr) \bigr|_{\z=q^{-1}\tau_j} \epp
\end{multline}
If we substitute $g(\z/\xi)=\Delta_{\z}^{-1}\psi(\z/\xi)$ and take
$\xi=\xi_2$, then we immediately arrive at the equality (\ref{equal}).

As was mentioned above, there is one more case to be considered,
corresponding to the contour $\G_0$, i.e.\ to the residue at $\z^2 = 0$
in equation (6.10) of \cite{JMS08pp} which has to vanish. Its vanishing
follows from
\begin{multline}
     \lim_{\x_1 \rightarrow 0} \x^{- \a} \bigl( \om (\x_1, \x_2) +
        \overline{D}_{\x_1} \overline{D}_{\x_2} \D^{-1}_{\x_1} \ps (\x)
        \bigr) = \\
        \frac{2 q^{- \k}}{q^\k + q^{- \k}} \biggl[
           \r(\x_2) - q^{- \a} + (q^\a - q^{- \a})
           \int_{\cal C} \rd m(\m) G(\m, \n_2) \biggr] = 0 \epp
\end{multline}
Here we have used (\ref{Psi}), (\ref{ompsi}), (\ref{defpsi}),
(\ref{actdbar}) as well as the fact that $\lim_{\n \rightarrow - \infty}
\r (\x) = (q^{\a + \k} + q^{- \a - \k})/(q^\k + q^{- \k})$ in the first
equation and the identity (\ref{grhoid}) in the second equation.

The normalization condition just shown to be satisfied by our function
$\om$ defined in (\ref{defom}) is the main ingredient in our proof
that $\om$ is in fact the same function as introduced in equation (7.2)
of \cite{JMS08pp}. Let us consider $\om$ as a function of $\x_1$. As
was shown in \cite{JMS08pp} the function $\r (\x_1)$ depends only on
$\x_1^2$. The same is then true for $\Psi (\x_1, \x_2)$ from (\ref{Psi}).
Using (\ref{ompsi}) we conclude that $\x^{- \a} \om(\x_1, \x_2|\k, \a)$
is a function of $\x_1^2$. From its definition (\ref{defom}) and from
(\ref{defdens}), (\ref{c2b11}) we see that $\om$ is rational in $\x_1^2$
of the form $P(\x_1^2)/Q(\x_1^2)$, where $P$ and $Q$ are polynomials.
Clearly both of them are at most of degree $N + 2$. The zeros of $Q$ are
the $N$ zeros of the transfer matrix eigenvalue $\La (\x_1, \k)$ plus
two zeros at $q^{\pm 2} \x_2^2$ stemming from the two simple poles of
$\x^{-\a} \cv^*_{[1,2]} (\x_2, \a) \bv^*_{[1,2]} (\x_1,\a - 1) (1)$.
Comparing now with the definition (7.2) of \cite{JMS08pp} we see that
the functions there has precisely the same structure. It is rational
of the form $\tilde P (\x_1^2)/ \tilde Q (\x_1^2)$ with two polynomials
$\tilde P$, $\tilde Q$ at most of degree $N + 2$. $Q$ and $\tilde Q$
have the same zeros. We may therefore assume that they are identical.
In order to show that $P$ and $\tilde P$ also agree we have to provide
$N + 3$ relations. $N + 1$ of them are given by the normalization
condition above. Another two come from the residues at the two trivial
poles.

Since they are outside the canonical contour, we have to consider
again the analytic continuation of the integral (\ref{Psi}) defining
$\Psi$ with respect to $\x_1$. There are four regions depending on the
\begin{figure}
    \centering
    \includegraphics{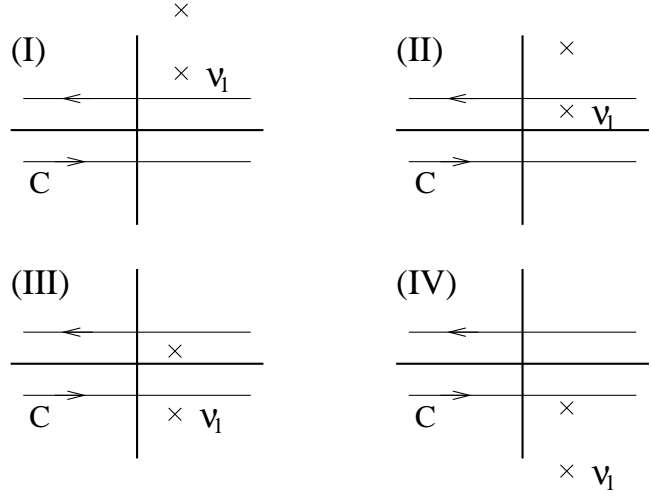}
    \caption{\label{fig:anacontpsi}
             Four cases to be considered for the analytic continuation
             of $\Psi (\x_1, \x_2)$ with respect to $\n_1$. Here $\cal C$
             is the canonical contour of figure~\ref{fig:cont}.}
\end{figure}
location of $\n_1$ relative to the contour (see figure
\ref{fig:anacontpsi}). Using (\ref{Psi}) we obtain
\begin{multline} \label{anacontpsi}
     \Ps (\x_1, \x_2) =
        \int_{\cal C} \rd m(\m) G(\m, \n_2)
	\bigl(q^\a \cth(\m - \n_1 - \h) - \r(\x_1) \cth(\m - \n_1) \bigr)
        \\[1ex] - \begin{cases}
           \dst{\frac{G(\n_1, \n_2)}{1 + \fa (\n_1, \k)}}
           & \text{case (I)} \\[2ex]
           0 & \text{case (II)} \\[1ex]
           \dst{\frac{G(\n_1, \n_2)}{1 + \fa (\n_1, \k)} +
                \frac{q^\a G(\n_1 + \h, \n_2)}
                     {(1 + \fa (\n_1 + \h, \k)) \r(q \x_1)}}
                 & \text{case (III)} \\[2ex]
           \dst{\frac{G(\n_1, \n_2)}{1 + \fa (\n_1, \k)}}
           & \text{case (IV)} \epp
        \end{cases}
\end{multline}
Then, e.g.\ be means of the integral equation (\ref{newg})
\begin{subequations}
\label{restrivpsi}
\begin{align} \label{resa}
     \res_{\x_1^2 = q^2 \x_2^2} & \Psi (\x_1, \x_2) = \notag \\
        & - \frac{2 \x_2^2 q^{2 - \a}}
               {(1 + \fa (\n_2 + \h, \k))(1 + \faq(\n_2, \k))} =
        - \frac{2 \x_2^2 q^{2 - \a} a(\x_2 q) d(\x_2)}
               {\La (\x_2 q, \k) \La(\x_2, \k)} \epc \\[2ex]
     \res_{\x_1^2 = q^{- 2} \x_2^2} & \Psi (\x_1, \x_2) = \notag \\
     \label{resb}
        & \frac{2 \x_2^2 q^{\a - 2}}
               {(1 + \fa (\n_2, \k))(1 + \faq(\n_2 - \h, \k))} =
        \frac{2 \x_2^2 q^{\a - 2} a(\x_2) d(\x_2 q^{- 1})}
               {\La (\x_2, \k) \La(\x_2 q^{-1}, \k)} \epc
\end{align}
\end{subequations}
where $a$ and $d$ are the vacuum expectation values of the diagonal
elements of $T(\z)$. Since the ratios on the right hand side are
invariant under changing the normalization of the $R$-matrix, we can
directly compare the residues obtained from (\ref{ompsi}),
(\ref{restrivpsi}) with those obtained from equation (7.2) of
\cite{JMS08pp}. We find agreement, which completes the proof.

What if we consider (\ref{ompsi}) as the definition of $\om$? Then,
in addition, we have to show that there are no poles other than the
two trivial ones and those at the location of the zeros of
$\La(\x_1, \k)$. But this is immediately clear from (\ref{anacontpsi}).
The integral has only poles at the zeros of $\La(\x_1, \k)$. In case
(I) there is one additional pole at $\n_1 = \n_2 + \h$ with residue
(\ref{resa}). The simple poles of $G (\n_1, \n_2)$ at $\la_j + \h$,
where the $\la_j$ are the Bethe roots (see appendix \ref{app:dermult}),
are canceled by the simple poles of $\fa (\n_1, \k)$ (see equation
(\ref{auxes})). In cases (II) and (IV) there is nothing to show. In
case (III) we have one additional pole at $\n_1 = \n_2 - \h$ with residue
(\ref{resb}). The simple poles at $\la_j - \h$ have vanishing
residue due to (\ref{auxes}) and since
\begin{equation}
     \res_{\n_1 = \la_j - \h} G(\n_1, \n_2) =
        - \frac{q^\a G(\la_j, \n_2)}{\r (\z_j) \fa' (\la_j)} \epp
\end{equation}

\section{The exponential form -- preliminary remarks}
\label{sec:top}
The main result of \cite{JMS08pp} is the formula (1.12). It makes
the calculation of arbitrary correlation functions possible, because
the operators $\tb^*,\bv^*,\cv^*$ generate a basis of the space of
quasi-local operators \cite{BJMS09app}. Although this formula proves the
factorization of the correlation functions and allows, in principle,
also for their direct numerical evaluation, it may be sometimes
preferable to avoid the creation operators and to have an explicit
formula for the correlation functions in the standard basis generated
by the local operators ${e_j}_\e^{\e'}$. We believe that some form of
the exponential formula discussed in the previous papers
\cite{BJMST08app,BGKS06,BGKS07,BDGKSW08} must be valid in case of
temperature, disorder and magnetic fields as well. Unfortunately, the
problem of constructing all operators that appear in this formula remains
still open. We hope to come back to it in a future publication. Here we
formulate the general properties we expect for these operators and show
by examples how they should look like for short distances.

From now on we shall use the notation and the terminology of the paper
\cite{BJMST08app}. In particular we shall be dealing with the space
${\cal W}^{(\a)}$ of quasi-local operators of the form $q^{2\a S(0)}\CO$
introduced there. First we define a density operator $D_N^{\ast}:
{\cal W}^{(\a)} \rightarrow {\mathbb C}$ which generalizes that one
defined by the formulae (33), (34) of \cite{BGKS07}, namely, for any
quasi-local operator $\CO$ we define
\begin{equation}
     D_N^\ast (\CO) = \< \CO \>_{T,\a,\kappa}
\end{equation}
in such a way that 
\begin{equation}
     D_N^\ast \bigl( {e_1}^{\e_1}_{\e_1'} \dots {e_m}^{\e_m}_{\e_m'}
        \bigr) =
    {D_N}^{\e_1' \dots \e_m'}_{\e_1 \dots \e_m}(\x_1, \dots, \x_m|\k, \a)
\end{equation}
where ${D_N}$ is the density matrix defined in (\ref{defdens}).

We expect that as before
\begin{equation} \label{exponential}
     D_N^\ast (\CO) = \mathbf{tr}^{\a} \bigl\{ \exp(\Omega)
                      \bigl( q^{2 \a S(0)} \mathcal{O} \bigr) \bigr\} \epc
\end{equation}
where $\mathbf{tr}^\a$ is the $\a$-trace defined in \cite{BJMST08app}
and where the operator $\Omega$ consists of two terms like that one
constructed in \cite{BGKS07}
\begin{equation} \label{Omega}
     \Omega = \Omega_1 + \Omega_2 \epp
\end{equation}
In fact, the first term follows from \cite{BJMST08app,JMS08pp}. It must
be of the form
\begin{equation} \label{Omega1}
     \Omega_1 = \int\frac{d\z_1^2}{2\pi i\z_1^2}
                \int\frac{d\z_2^2}{2\pi i\z_2^2}
                \bigl(\om_0(\z_1/\z_2|\a) - \om(\z_1,\z_2|\k,\a) \bigr)
                \bv(\z_1)\cv(\z_2) \epc
\end{equation}
where the function $\om_0$ was defined in \cite{BJMST08app},
\begin{equation} \label{om0}
     \om_0(\z|\a) = - \biggl( \frac{1-q^{\a}}{1+q^{\a}} \biggr)^2
                      \Delta_{\z} \psi(\z) \epp
\end{equation}
The second part in the right hand side of (\ref{Omega}) should be of the
form 
\begin{equation}
     \Omega_2 = \int\frac{d\z^2}{2\pi i\z^2}\log(\rho(\z))\tb(\z)
\label{Omega2}
\end{equation}
where the operator $\tb$ is yet to be determined. In some sense
it must be the conjugate of the operator $\tb^*$. The integration
contour for both, $\Om_1$ and $\Om_2$, is taken around all simple
poles $\z_1,\z_2,\z=\xi_j$ with $j=1,\dots,m$ in anti-clockwise
direction. The number $m$ is the length of locality of the operator
$\CO$.

Let us list some of the most important expected properties of the
operator $\tb$. First, we expect that like $\tb ^*(\z )$ the operator
$\tb(\z)$  is block diagonal,
\[
     \tb(\z ):\ \ \mathcal{W}_{\a ,s} \rightarrow \mathcal{W}_{\a ,s} \epc
\]
where, as was explained in \cite{BJMST08app}, $\mathcal{W}_{\a ,s}
\subset \mathcal{W}^{(\a)}$ is the space of quasi-local operators of
spin $s$. We will deal below mostly with the sector $s=0$.

Then we expect $\tb(\z)$ to have simple poles at $\z=\xi_j$. Let us define
\begin{equation} \label{tbj}
     \tb_j = \res_{\z=\xi_j} \tb(\z) \frac{d\z^2}{\z^2}
\end{equation}
while
\begin{equation}
     \tb^*_j = \tb^*(\xi_j) \epp
\label{tb*j}
\end{equation}
In contrast to (\ref{tbj}) the operator $\tb^*_j$ is well defined only
if it acts on the states $X_{[k,l]}$ with $l<j$. This will be always
implied below. Let us denote $\tb_{[k,l]}(\z)$ and respectively
$\tb_{j[k,l]}$ the operators defined on the interval $[k,l]$ with 
$k\le j\le l$.

We also expect that $R$-matrix symmetry holds similar to the formula
(2.16) of \cite{BJMST08app} for $\tb^*$,
\begin{equation} \label{Rsymmetry}
     \sb_i\,\tb_{[k,l]}(\z)=\tb_{[k,l]}(\z)\,
                            \sb_i\quad\mbox{for}\quad k\le i<l \epp
\end{equation}
Here as was defined in \cite{BJMST08app}
\begin{subequations}
\begin{align}
&\sb_i=K_{i,i+1}\check \bR_{i,i+1}(\xi_i/\xi_{i+1}) \epc \\[1ex]
&\check\bR_{i,i+1}(\xi_i/\xi_{i+1})(X)=\check R_{i,i+1}(\xi_i/\xi_{i+1})X
\check R_{i,i+1}(\xi_i/\xi_{i+1})^{-1} \epc \\[1ex]
&\check R_{i,i+1}(\z)=P_{i,i+1}R_{i,i+1}(\z) \epc
\end{align}
\end{subequations}
where $K_{i,j}$ stands for the transposition of arguments $\xi_i$ and
$\xi_j$ and $P_{i,j}\in\End(V_i\otimes V_j)$ is the transposition matrix.

The further properties are:
\begin{itemize}
\item
{\it Commutation relations }
\begin{equation} \label{comrel}
     [\tb_j,\tb_k]_- = [\tb_j,\bv(\z_1)\cv(\z_2)]_- = 0 \epp
\end{equation}
\item
{\it Projector property}
\begin{equation} \label{projector}
     \tb_j^2 = \tb_j \epp
\end{equation}
\item
{\it Relations with $\tb^*$}
\begin{align} \label{tbtb*}
     &\tb^{}_j\tb^*_k=\tb^*_j\tb^{}_k\quad
      \mbox{for}\quad j\ne k \epc \notag \\[1ex]
     &\tb^{}_j\tb^*_{j} = \tb^*_{j},\quad\tb^*_{j}\tb^{}_j = 0 \epp
\end{align}
\item
{\it Reduction properties}
\begin{subequations}
\label{redt}
\begin{align}
     &\tb_{1[1,l]}(q^{\a\s^z_1}X_{[2,l]})
        =q^{\a\s^z_1}X_{[2,l]} \label{redrighttb1} \epc \\[1ex]
     &\tb_{j[1,l]}(q^{\a\s^z_1}X_{[2,l]})
        =q^{\a\s^z_1}\tb_{j[2,l]}(X_{[2,l]})\quad\mbox{for} \quad 1<j\le l
         \label{redrighttbj} \epc \\[1ex]
     &\tb_{j[1,l]}(X_{[1,l-1]})
        =\tb_{j[1,l-1]}(X_{[1,l-1]})\quad\mbox{for} \quad 1\le j< l 
         \label{redlefttbj} \epc \\[1ex]
     &\tb_{l[1,l]}(X_{[1,l-1]})=0 \epp \label{redlefttbl}
\end{align}
\end{subequations}
\end{itemize}

Let us comment on these relations. First of all the commutation
relations (\ref{comrel}) lead to the factorization of the exponential
\begin{equation} \label{factoromega}
     \exp(\Omega)=\exp(\Omega_1+\Omega_2)=\exp(\Omega_1)\exp(\Omega_2) \epp
\end{equation}
As we know (see \cite{BJMST08app,BGKS06,BGKS07} and earlier references
therein) the operator $\Omega_1$ becomes nilpotent when it acts on
states of finite length. By way of contrast, the operator $\Omega_2$
is not nilpotent, but due to (\ref{comrel}) and the projector property
(\ref{projector}) one can conclude that
\begin{equation} \label{Omega2viat}
     \exp(\Omega_2)\bigl(q^{2\a S(0)}X_{[1,l]}\bigr)=
        \prod_{j=1}^l (1-\tb_{j[1,l]}+\rho_j \tb_{j[1,l]})
                      \bigr(X_{[1,l]}\bigr) q^{2\a S(0)} \epc
\end{equation}
where $\rho_j=\rho(\xi_j)$. The reduction properties (\ref{redt}) look
standard except for the first one. It is easy to see that we need all 
of them in order to have the reduction property (\ref{reductionD}) of
the density matrix, but still we do not have a good understanding of
(\ref{redrighttb1}).

Let us show how the $\tb_j$ for $s=0$ explicitly look like in two
particular cases, namely, for $m=1$ and $m=2$ where $m=l-k+1$
and without loss of generality $k=1$.


\noindent $\mathbf{m=1:}$
\begin{equation}
\tb_{1[1,1]}=\frac{q^{\a\s^z_1}\otimes\s^z_1}{q^{\a}-q^{-\a}}
\label{tb1n1}
\end{equation}

\noindent $\mathbf{m=2:}$
\begin{align} \label{tb1n2}
     \tb_{1[1,2]}&=\frac{1}{4}\;\frac{q^{\a}+q^{-\a}}{q^{\a}-q^{-\a}}
        I\otimes\biggl[\s^z_1- \frac{q-q^{-1}}{\xi_1/\xi_2-\xi_2/\xi_1}
        \cdot(\s_1^+\s_2^- - \s_1^-\s_2^+)\biggr]+
        \notag\\[1ex]
     &+\frac{1}{4}\s^z_1\otimes\biggl[\s^z_1-
       \frac{q-q^{-1}}{\xi_1/\xi_2-\xi_2/\xi_1}
        \cdot(\s_1^+\s_2^- - \s_1^-\s_2^+)\biggr]+
        \notag\\[1ex]
     &+\frac{1}{4}q^{\a\s^z_1}\s^z_2\otimes\biggl[\biggl(
       \frac{q^{\s^z_1}}{q^{\a+1}-q^{-\a-1}}+
       \frac{q^{-\s^z_1}}{q^{\a-1}-q^{-\a+1}}\biggr)\s^z_1\s^z_2-
       \notag\\[1ex]
     &-\frac{(q-q^{-1})(q^{\a}+q^{-\a})}
            {2(q^{\a+1}-q^{-\a-1})(q^{\a-1}-q^{-\a+1})}
       (\xi_1/\xi_2-\xi_2/\xi_1)\cdot(\s_1^+\s_2^- - \s_1^-\s_2^+)-
       \notag\\[1ex]
     &-\frac{(q+q^{-1})(q^{\a}-q^{-\a})}
            {2(q^{\a+1}-q^{-\a-1})(q^{\a-1}-q^{-\a+1})}
       (\xi_1/\xi_2+\xi_2/\xi_1)\cdot(\s_1^+\s_2^- + \s_1^-\s_2^+)\biggr]
       \epp
\end{align}
The operator $\tb_{2[1,2]}$ can be obtained from $\tb_{1[1,2]}$ using the
$R$-matrix symmetry (\ref{Rsymmetry}),
\begin{equation} \label{tb2n2}
     \tb_{2[1,2]}=\sb_1\tb_{1[1,2]}\sb_1^{-1} \epp
\end{equation}
One can check that all the above properties are fulfilled. 

It is interesting to understand how the limit $\a\rightarrow 0$ works,
because, as one can see from (\ref{tb1n1})-(\ref{tb2n2}), the operators
$\tb_{1}$ and $\tb_{2}$ are singular in this limit. More precisely, only
the very first term in the expression (\ref{tb1n2}) for $\tb_{1[1,2]}$
is singular. In fact, they contribute into the density matrix only in
such a combination that this singularity cancels. So, we actually need
to calculate the residue with respect to $\a$. In appendix \ref{app:alpha}
we will discuss this issue in more detail. We will also show how these
operators are related to the `fermionic' operators $\mathbf{h}_j$
constructed in \cite{BGKS07,BDGKSW08}.

\section{Conclusions}
The main result of this work is the description of the function
$\om$ in terms of integrals involving the auxiliary function
$\fa$ and the function $G$, which are solutions of integral equations.
As we have experienced in our previous work such type of description is
efficient for performing the Trotter limit and for the actual
numerical evaluation of correlation functions. We know from
\cite{JMS08pp} that no other functions than $\om$ and $\r$ are required.
Thus, together with \cite{JMS08pp}, we have achieved a rather complete
understanding of the mathematical structure of the static correlation
functions of the XXZ chain. Our results are equally valid in the finite
temperature as in the finite length case, the only difference being
a different driving term in the non-linear integral equation for
the auxiliary functions $\fa$.

We expect that our results open a way
for further concrete studies of short range correlation functions
at finite temperature as initiated in \cite{BGKS06,BGKS07,BDGKSW08}.
We hope that in the future it will also prove useful in studying
field theoretical scaling limits as well as the large distance
asymptotics of correlation functions in the XXZ chain.

We have obtained our expression for $\om$ through a novel multiple
integral representation of the density matrix of the XXZ chain
including a disorder parameter $\a$. We think that this multiple
integral representation is also interesting on its own right.

We would like to point out that we obtained a remarkably beautiful
and simple characterization of the function $\Psi$ on the inhomogeneous
finite lattice through equation (\ref{eqPsi1}) and the
residua (\ref{restrivpsi}). The function $\Psi$ is important because
it becomes the transcendental part of $\om$ in the Trotter limit.

We further performed a case study, looking for an operator $\tv$,
adjoint to $\tv^*$ that allows us to write the density matrix
in an exponential form even in the presence of a disorder parameter
and a finite magnetic field. We obtained explicit expressions
for $\tv$ for $m = 1, 2$, but so far could not find a general construction
behind it. We hope to come back to this latter point in the future.
\\[1ex]
{\bf Acknowledgement.}
The authors are deeply indebted to M.~Jimbo, T.~Miwa and F.~Smirnov
for a fruitful collaboration that led to this paper. They would like to
thank A.~Kl\"umper, K. Nirov, J. Suzuki, C. Trippe and A. Wei{\ss}e for
helpful and stimulating discussions. Special thanks are due to
N.~Kitanine for explaining how to derive the multiple integral
representation in section~\ref{sec:multint}. This work was financially
supported by the Volkswagen Foundation.


{\appendix
\Appendix{Derivation of the multiple integral representation}
\label{app:dermult}
\noindent
A multiple integral representation for the density matrix
in the non-twisted case $\a = 0$ was derived in \cite{GHS05}.
Here we shall only indicate which modifications are necessary to
include non-zero $\a$ and otherwise refer the reader to that work.

First note that
\begin{equation} \label{densdens}
	\frac{\<\k + \a| T^{\e_1'}_{\e_1} (\x_1, \k) \dots
	                 T^{\e_m'}_{\e_m} (\x_m, \k) |\k\>}
             {\<\k + \a|\prod_{j=1}^m t (\x_j,\k)|\k\>} =
	\frac{\<\k| T^{\e_m}_{\e_m'} (\x_m, \k) \dots
	                 T^{\e_1}_{\e_1'} (\x_1, \k) |\k + \a\>}
             {\<\k|\prod_{j=1}^m t (\x_j,\k)|\k + \a\>} \epc
\end{equation}
because of the symmetry of the $R$-matrix with respect to transposition.
We may therefore start our calculation with
\begin{equation} \label{densapp}
	\frac{\<\k| T^{\a_1}_{\be_1} (\x_1, \k) \dots
	                 T^{\a_m}_{\be_m} (\x_m, \k) |\k + \a\>}
             {\<\k|\prod_{j=1}^m t (\x_j,\k)|\k + \a\>} \epc
\end{equation}
which brings us closer to the notation of \cite{GHS05}.

The left and right eigenvectors $\<\k|$ and $|\k + \a\>$ can be
constructed by means of the algebraic Bethe ansatz. They are
parameterized by two sets $\{\la\} = \{\la_j\}_{j=1}^{N/2}$ and $\{\m\} =
\{\m_j\}_{j=1}^{N/2}$ of Bethe roots, which are special solutions
to the Bethe ansatz equations
\begin{equation} \label{baela}
     \frac{q^{- 2 \k} d(\la_j)}{a(\la_j)} \prod_{k=1}^{N/2}
        \frac{\sh(\la_j - \la_k + \h)} {\sh(\la_j - \la_k - \h)} = - 1
        \epc \qd
     \frac{q^{- 2 \k - 2 \a} d(\m_j)}{a(\m_j)} \prod_{k=1}^{N/2}
         \frac{\sh(\m_j - \m_k + \h)} {\sh(\m_j - \m_k - \h)} = - 1
\end{equation}
for $j = 1, \dots, N/2$. By $a(\la)$ and $d(\la)$ we denoted here the
vacuum expectation values of the diagonal elements of $T(\z)$. 
For its $\k$-twisted version we shall reserve the notation
\begin{equation}
     T(\z|\k) = \begin{pmatrix} A(\la) & B(\la) \\ C(\la) & D(\la)
                \end{pmatrix} \epp
\end{equation}
for the matrix elements. Then the eigenvectors $\<\k|$ and $|\k + \a\>$
are
\begin{subequations}
\begin{align}
     \<\k| = \<\{\la\}| & = \<0| C(\la_1) \dots C(\la_{N/2}) \epc \\[1ex]
     |\k + \a\> = |\{\m\}\> & = B(\m_1) \dots B(\m_{N/2}) |0\> \epc
\end{align}
\end{subequations}
where $\<0|$ and $|0\>$ are the left and right pseudo vacuum states.

With the solutions $\{\la\}$ and $\{\m\}$ of the Bethe ansatz equations
we associate the auxiliary functions
\begin{subequations}
\begin{align} \label{auxes}
     \fa (\la) & =
     \fa(\la,\k) = \frac{q^{- 2 \k} d(\la)}{a(\la)}
                   \prod_{k=1}^{N/2} \frac{\sh(\la - \la_k + \h)}
		                          {\sh(\la - \la_k - \h)} \epc \\
     \fa_\a (\la) & = 
     \fa(\la,\k + \a) = \frac{q^{- 2 \k - 2 \a} d(\la)}{a(\la)}
                   \prod_{k=1}^{N/2} \frac{\sh(\la - \m_k + \h)}
		                          {\sh(\la - \m_k - \h)}
\end{align}
\end{subequations}
and the ratio of $q$-functions
\begin{equation}
     \Ph (\la) = \prod_{j=1}^{N/2} \frac{\sh(\la - \m_j)}{\sh(\la - \la_j)}
                 \epp
\end{equation}
Then $\<\{\la\}|$ is the `dominant' left eigenvector of $t(\z,\k)$ with
eigenvalue
\begin{equation} \label{evla}
     \La (\z,\k) =
         q^{\k} a(\la) \biggl[ \prod_{j=1}^{N/2}
         \frac{\sh(\la - \la_j - \h)}{\sh(\la - \la_j)} \biggr]
         \bigl(1 + \fa (\la) \bigr) \epc
\end{equation}
and similarly $|\{\m\}\>$ is the dominant right eigenvector of the
$\alpha$-twisted transfer matrix $t (\z,\k + \a)$ with eigenvalue
\begin{equation} \label{evmu}
     \La (\z, \k + \a) =
         q^{\k + \a} a(\la) \biggl[ \prod_{j=1}^{N/2}
         \frac{\sh(\la - \m_j - \h)}{\sh(\la - \m_j)} \biggr]
         \bigl(1 + \fa_\a (\la) \bigr)
\end{equation}
Dividing (\ref{evmu}) by (\ref{evla}) we obtain the identities
\begin{equation} \label{rhophi}
     \r (\z) = \frac{1 + \fa_\a (\la)}{1 + \fa (\la)}
                q^\a \Ph (\la - \h) \Ph^{-1} (\la)
              = \frac{1 + \faq_\a (\la)}{1 + \faq (\la)}
                q^{- \a} \Ph (\la + \h) \Ph^{-1} (\la)
\end{equation}
which will be needed below. Here we have introduced the notation
$\faq = 1/\fa$ and $\faq_\a = 1/\fa_\a$. In the derivation of the
multiple integral formula below we shall use that $\r$ is analytic
and non-zero inside the canonical contour $\cal C$ which follows
from the Bethe equations and from the explicit form of the vacuum
expectation values $a$ and $d$.

The derivation of the density matrix for $\alpha = 0$ in \cite{GHS05}
is divided into two steps. Step 1 is the derivation of the `general
left action' $\<\{\la\}|T^{\a_1}_{\be_1} (\x_1,\k) \dots
T^{\a_m}_{\be_m} (\x_m,\k)$ of a string of monodromy matrix elements on
the left dominant eigenvector. This step remains the same as before. The
general left action is given by Lemma 1 of \cite{GHS05}. In a second
step one must calculate ratios of scalar products of the form
\begin{equation} \label{defchi}
     \chi = \frac{\<\{\n^+\} \cup \{\la^-\}|\{\m\}\>}
          {\<\{\la\}|\{\m\}\>\prod_{j=1}^m \La (\x_j, \k)}
\end{equation}
which are generated in step 1. A useful expression for these ratios
in the untwisted case $\a = 0$ is provided by lemma 2 of \cite{GHS05}.
This needs to be modified here.

The notation in (\ref{defchi}) is meant as follows. We divide the
sets $\{\la\}$ and $\{\n\} = \{\n_j\}_{j=1}^m = \{\ln \x_j\}_{j=1}^m$
into disjoint subsets $\{\la^+\}$, $\{\la^-\}$ and $\{\n^+\}$,
$\{\n^-\}$, such that their unions are $\{\la^+\} \cup \{\la^-\}
= \{\la\}$ and $\{\n^+\} \cup \{\n^-\} = \{\n\}$. The number of
elements in a set $\{x\}$ will be denoted $|x|$. We shall assume that
$|\la^+| = |\n^+| = n$. For a given partition of $\{\la\}$ into
$\{\la^+\}$, $\{\la^-\}$ we order the $\la$s such that
\begin{equation}
     (\la_1, \dots, \la_{N/2}) =
        (\la_1^+, \dots, \la_n^+, \la_1^-, \dots, \la_{N/2-n}^-)
\end{equation}
and we define
\begin{equation}
     (\tilde \la_1, \dots, \tilde \la_{N/2}) =
        (\n_1^+, \dots, \n_n^+, \la_1^-, \dots, \la_{N/2-n}^-) \epp
\end{equation}
Then using lemma 2 of \cite{GKS04a} (with the roles of $\la$ and $\m$
interchanged) we arrive after some trivial cancellations at
\begin{multline} \label{ratio}
     \chi = \Biggl[ \prod_{j=1}^{|\n^-|}
	          \frac{\prod_{k = 1}^{N/2} b(\la_k - \n_j^-)}
		       {a(\n_j^-)(1 + \fa (\n_j^-))} \Biggr]
          \Biggl[ \prod_{j=1}^{|\la^+|}
	          \frac{\prod_{k=1, \la_k \ne \la_j^+}^{N/2}
		        b(\la_k - \la_j^+)}
		       {a(\la_j^+) (1 + \fa (\n_j^+))} \Biggr]
	       \tst{\det^{-1}_{|\la^+|}} \biggl[\frac{c(\la_j^+ - \n_k^+)}
	                        {b(\la_j^+ - \n_k^+)}\biggr] \\ \cdot
          \underbrace{
          \frac{\det_{N/2} \widehat N (\m_j, \tilde \la_k)}
               {\det_{N/2} \widehat N (\m_j, \la_k)}
          \biggl[ \prod_{j=1}^{|\la^+|}
          \Ph (\n_j^+ - \h) \Ph^{-1} (\la_j^+ - \h) \biggr]
          }_{\tst{= \Xi}} \epc
\end{multline}
where
\begin{equation}
     \widehat N (\m_j, \la_k)
         = \re (\m_j - \la_k) - \re (\la_k - \m_j) \fa_\a (\la_k) \epp
\end{equation}

In the following we concentrate on the term $\Xi$ in the second
line of (\ref{ratio}). We want to transform it into a form that
allows us to perform the Trotter limit $N \rightarrow \infty$.
We define column vectors $\uv_k$, $k = 1, \dots, |\n^+| = n$, and
$\vv_k$, $k = 1, \dots, N/2$, by
\begin{align}
     & (\uv_k)^j = \Ph (\n_k^+ - \h) \widehat N (\m_j, \n_k^+)
                 = \Ph (\n_k^+ - \h) \bigl(
                   \re (\m_j - \n_k^+) - \re (\n_k^+ - \m_j)
		   \fa_\a (\n_k^+) \bigr)
       \epc \notag \\[1ex] \label{defvv}
     & (\vv_k)^j = \Ph (\la_k - \h) \widehat N (\m_j, \la_k)
                 = \re (\m_j - \la_k) \Ph (\la_k - \h) + q^{- 2 \a}
                   \re (\la_k - \m_j) \Ph (\la_k + \h) \epp
\end{align}
In (\ref{defvv}) we used the Bethe ansatz equations (\ref{baela})
to eliminate $q^{- 2 \k} d(\la_k)/a(\la_k)$. We have
\begin{equation}
     \Xi = \frac{\det (\uv_1, \dots, \uv_n, \vv_{n+1}, \dots, \vv_{N/2})}
                {\det (\vv_1, \dots, \vv_{N/2})} \epp
\end{equation}

Next we use a trick we learned from N. Kitanine \cite{Kitanine08up} and
which in a similar form originally appeared in \cite{IKMT99}. Define
a matrix $X$ with matrix elements
\begin{equation}
     X^j_l = q^{\a} \cth (\m_l - \la_j) \res_{\la = \m_l} \Ph^{-1} (\la)
\end{equation}
and column vectors $\Uv_k = \Uv (\n_k^+) = X \uv_k$, and $\Vv_k =
X \vv_k$. Then
\begin{equation}
     \Xi = \frac{\det (\Uv_1, \dots, \Uv_n, \Vv_{n+1}, \dots, \Vv_{N/2})}
                {\det (\Vv_1, \dots, \Vv_{N/2})} \epp
\end{equation}
The vectors $\Uv_k$ and $\Vv_k$ are easily calculated by means of the
residue theorem. Choose two simple closed contours ${\cal C}_\la$ and
${\cal C}_\m$ such that all Bethe roots $\la_j$ are inside ${\cal C}_\la$
but outside ${\cal C}_\m$ and all Bethe roots $\m_j$ are inside
${\cal C}_\m$ but outside ${\cal C}_\la$. Then
\begin{multline}
     (\Vv_k)^j =  \\ \sum_{l=1}^{N/2} q^{\a} \cth(\m_l - \la_j)
                 \res_{\la = \m_l} \Ph^{-1} (\la)
                 \bigl[ \re (\m_l - \la_k) \Ph (\la_k - \h)
                        + q^{- 2 \a} \re (\la_k - \m_l) \Ph (\la_k + \h)
			\bigr]
                 \\
               = \int_{{\cal C}_\m} \frac{\rd \m}{2 \p \i}
                 \underbrace{
                 \cth(\m - \la_j) \Ph^{-1} (\m)
                 \bigl[q^{\a} \re (\m - \la_k) \Ph (\la_k - \h)
                        + q^{- \a} \re (\la_k - \m) \Ph (\la_k + \h)
			\bigr]
                 }_{\tst{= f(\m)}} \epp
\end{multline}
The function $f$ is periodic with period $\i \p$, and
$\lim_{\Re \m \rightarrow \pm \infty} f(\m) = 0$. Moreover,
$\Ph^{-1} (\m)$ is analytic inside ${\cal C}_\la$. Hence, for $j \ne k$,
\begin{multline}
     (\Vv_k)^j = - \int_{{\cal C}_\la} \frac{\rd \m}{2 \p \i} f(\m)
               = - (\res_{\m = \la_j} + \res_{\m = \la_k} +
                    \res_{\m = \la_k + \h} + \res_{\m = \la_k - \h})
                   f(\m) \\
               = q^{- \a} \cth(\la_j - \la_k - \h) -
                 q^{\a} \cth(\la_j - \la_k + \h) = K_\a (\la_j - \la_k)
                 \epp
\end{multline}
For $j = k$ we have a non-trivial residue from the second order pole
at $\la_j$, and
\begin{equation} \label{vvdiag}
     (\Vv_j)^j = (q^{- \a} \Ph (\la_j + \h) - q^{\a} \Ph (\la_j - \h))
                 \6_\m \Ph^{- 1} (\m) \bigr|_{\m = \la_j}
                 + K_\a (0) \epp
\end{equation}
Similarly, we obtain
\begin{multline} \label{uv}
     (\Uv_k)^j = q^{\a} \cth(\la_j - \n_k^+)
                 \Ph^{-1} (\n_k^+) \Ph (\n_k^+ - \h)
                 \bigl[ 1 + \fa_\a (\n_k^+) \bigr]
                 - q^{\a} \cth(\la_j - \n_k^+ + \h) \\[1ex]
                 - q^{\a} \cth(\la_j - \n_k^+ - \h)
                 \Ph^{-1} (\n_k^+ + \h) \Ph (\n_k^+ - \h) \fa_\a (\n_k^+)
                 \epp
\end{multline}

Following once more \cite{Kitanine08up} we eliminate the function
$\Ph$ from (\ref{vvdiag}) and (\ref{uv}). This can be done by means
of the identities (\ref{rhophi}). For the first term on the right hand
side of (\ref{vvdiag}) we obtain
\begin{multline}
     \bigl(q^{- \a} \Ph (\la_j + \h) - q^{\a} \Ph (\la_j - \h)\bigr)
         \6_\m \Ph^{- 1} (\m) \bigr|_{\m = \la_j} = \\
         \lim_{\la \rightarrow \la_j}
         \bigl(q^{- \a} \Ph (\la + \h) - q^{\a} \Ph (\la - \h)\bigr)
         \frac{\fa'(\la) \Ph^{- 1} (\la)}{1 + \fa(\la)} = \\
         \lim_{\la \rightarrow \la_j} \fa' (\la) \r(\z)
         \biggl[ \frac{\faq (\la)}{1 + \faq_\a (\la)} -
                 \frac{1}{1 + \fa_\a (\la)} \biggr]
         = - \fa' (\la_j) \r(\z_j) \epc
\end{multline}
where we used (\ref{rhophi}) in the second equation and the Bethe
ansatz equation $\faq(\la_j) = - 1$ in the third equation. It follows
that
\begin{equation}
     (\Vv_k)^j = - \de^j_k \fa' (\la_j) \r(\z_j) + K_\a (\la_j - \la_k)
                 \epp
\end{equation}
Eliminating $\Ph$ from (\ref{uv}) by means of (\ref{rhophi}) we end up
with
\begin{multline}
     (\Uv_k)^j = \cth(\la_j - \n_k^+) (1 + \fa(\n_k^+)) \r(\x_k^+)
                 \\[1ex]
                 - q^{\a} \cth(\la_j - \n_k^+ + \h)
                 - q^{- \a} \cth(\la_j - \n_k^+ - \h) \fa (\n_k^+) \epp
\end{multline}

From here on we can proceed as in \cite{GKS04a}. Define a matrix
$V = (\Vv_1, \dots, \Vv_{N/2})$ and a vector $\Wv (\n) = V^{-1} \Uv (\n)$.
Then
\begin{multline} \label{psiisdetw}
     \Xi = \tst{\det_{N/2}}
           \bigl( V^{-1} \Uv_1, \dots, V^{-1} \Uv_n, \ev_{n+1},
                  \dots, \ev_{N/2} \bigr)
         = \\ \tst{\det_n} \bigl( \< \ev_j, V^{-1} \Uv_k \> \bigr)
         = \tst{\det_n} \bigl( \Wv (\n_k^+)^j \bigr) \epp
\end{multline}
$\Wv (\n_k^+)$ is the solution of the linear equation $V \Wv (\n_k^+)
= \Uv (\n_k^+)$, or, explicitly,
\begin{multline}
     \cth(\la_j - \n_k^+) (1 + \fa(\n_k^+)) \r(\x_k^+)
        - q^{\a} \cth(\la_j - \n_k^+ + \h)
        - q^{- \a} \cth(\la_j - \n_k^+ - \h) \fa (\n_k^+) \\
      = - \Wv (\n_k^+)^j \fa' (\la_j) \r(\z_j)
        + \sum_{l=1}^{N/2} K_\a (\la_j - \la_l) \Wv (\n_k^+)^l \epp
\end{multline}

This can be transformed into a linear integral equation. For this
purpose define
\begin{multline}
     G(\la, \n)
        = \frac{q^{\a} \cth(\la - \n + \h)}{1 + \fa (\n)}
        + \frac{q^{- \a} \cth(\la - \n - \h)}{1 + \faq (\n)} \\
        - \cth(\la - \n) \r(\x)
        + \sum_{l=1}^{N/2} \frac{K_\a (\la - \la_l) \Wv (\n)^l}
                            {1 + \fa (\n)} \epp
\end{multline}
This function is defined such that
\begin{equation} \label{gw}
     G(\la_j, \n_k^+)
        = \frac{\r (\z_j) \fa'(\la_j) \Wv (\n_k^+)^j}{1 + \fa(\n_k^+)}
          \epp
\end{equation}
We shall assume that $\n$ is located inside the canonical contour $\cal C$
shown in figure~\ref{fig:cont}. By construction $G(\la, \n)$ is then
meromorphic inside $\cal C$ and has a single simple pole with residue
$- \r(\x)$ at $\la = \n$. Using (\ref{gw}) it follows that
\begin{multline}
     G(\la, \n_k^+) = q^{- \a} \cth(\la - \n_k^+ -\h)
        - \cth(\la - \n_k^+) \r(\x_k^+) \\
        + \int_{\cal C} \frac{\rd \m}{2 \p \i}
          \frac{G(\m, \n_k^+) K_\a (\la - \m)}
               {\r (\re^\m) (1 + \fa (\m))}
\end{multline}
which is a linear integral equation for $G$.

Combining (\ref{psiisdetw}) and (\ref{gw}) we infer that
\begin{equation}
     \Xi = \biggl[ \prod_{j=1}^{|\la^+|}
                   \frac{1 + \fa (\n_j^+)}{\fa' (\la_j^+) \r (\z_j^+)}
                   \biggr] \tst{\det_{|\la^+|}}
                           \bigl( G (\la_j^+, \n_k^+) \bigr)
			   \epp
\end{equation}
Inserting this into (\ref{ratio}) we arrive at
\begin{equation} \label{lemtwosubst}
     \chi = \Biggl[ \prod_{j=1}^{|\n^-|}
	          \frac{\prod_{k = 1}^{N/2} b(\la_k - \n_j^-)}
		       {a(\n_j^-)(1 + \fa (\n_j^-))} \Biggr]
          \Biggl[ \prod_{j=1}^{|\la^+|}
	          \frac{\prod_{k=1, \la_k \ne \la_j^+}^{N/2}
		        b(\la_k - \la_j^+)}
		       {a(\la_j^+) \fa' (\la_j^+) \r (\z_j^+)} \Biggr]
                  \frac{\tst{\det_{|\la^+|}}
                        \bigl( G (\la_j^+, \n_k^+) \bigr)} 
	               {\tst{\det_{|\la^+|}}
                        \Bigl[\frac{c(\la_j^+ - \n_k^+)}
	                       {b(\la_j^+ - \n_k^+)}\Bigr]} \epp
\end{equation}
This replaces the expression in lemma 2 of \cite{GHS05}. Comparing
the two expressions we see only one explicit difference which is the
appearance of the additional factor $\r(\z_j^+)$ in the denominator.
It always comes with a factor $\fa' (\la_j^+)$. Therefore it is
easy to trace the modification required in the derivation of the multiple
integral formula in \cite{GHS05}. There is, however, also an implicit
change in the formula. The residue of $G(\la,\n)$ at $\la = \n$
is $- \r(\x)$ instead of $-1$. This causes another tiny modification in
the derivation of the multiple integral formula. In the first equation
(65) of \cite{GHS05} additional $m - |\la^+|$ factors of $\r^{-1}$
appear which together with the $|\la^+|$ explicit factors gives
exactly one factor per integral. The final result for the density
matrix matrix is
\begin{multline} \label{dnull}
     \frac{\<\k| T^{\a_1}_{\be_1} (\x_1, \k) \dots
                 T^{\a_m}_{\be_m} (\x_m, \k) |\k + \a\>}
          {\<\k|\prod_{j=1}^m t (\x_j,\k)|\k + \a\>} \\
     = \biggl[ \prod_{j=1}^{|\a^+|}
               \int_{\cal C} \frac{\rd \m_j}{2 \p \i}
               \frac{F_j (\m_j)}{\r (\re^{\m_j}) (1 + \fa (\m_j))} \biggr]
	       \mspace{-3.mu}
       \biggl[ \prod_{j = |\a^+| + 1}^m
               \int_{\cal C} \frac{\rd \m_j}{2 \p \i}
               \frac{\overline{F}_j (\m_j)}
                    {\r (\re^{\m_j}) (1 + \faq (\m_j))} \biggr] \\
       \frac{\det[ - G(\m_j, \n_k)]}
            {\dst{\prod_{1 \le j < k \le m}} \sh(\m_j - \m_k - \h)
                  \sh(\n_k - \n_j)} \epc
\end{multline}
where we have used the notation of \cite{GHS05}, i.e.
\begin{subequations}
\label{defffbar}
\begin{align}
     F_j (\la) & = \prod_{k=1}^{x_j - 1} \sh(\la - \n_k - \h)
                   \prod_{k = x_j + 1}^m \sh(\la - \n_k) \epc
		   \qd j = 1, \dots, |\a^+| \epc \\ 
     \overline{F}_j (\la) & =
                   \prod_{k=1}^{x_j - 1} \sh(\la - \n_k + \h)
		   \prod_{k = x_j + 1}^m \sh(\la - \n_k) \epc
		   \qd j = |\a^+| + 1, \dots, m \epc
\end{align}
\end{subequations}
and for $j = 1, \dots, |\a^+|$ we define $x_j$ to be the position
of the $(|\a^+| - j + 1)$th `$+$' in the sequence of upper indices
$(\a_j)_{j=1}^m$ while for $j = |\a^+| + 1, \dots, m$ it means the
position of the $(j - |\a^+|)$th `$-$' in the sequence of lower indices
$(\be_j)_{j=1}^m$.

Finally we replace $\x_k$ by $\x_{m - k + 1}$ and define $\eps_{k} =
\a_{m - k + 1}$, $\eps_{k}' = \be_{m - k + 1}$ for $k = 1, \dots, m$.
We denote the position of the $j$th `$+$' in $(\e_j)_{j=1}^m$ by
$\e_j^+$ , the position of the $j$th `$-$' in $(\e_j')_{j=1}^m$ by
$\e_j^-$. Then
\begin{equation}
     x_j = \begin{cases} m - \e_j^+ + 1 & j = 1, \dots, p \\
                         m - \e_{m - j + 1}^- + 1 & j = p + 1, \dots, m
           \end{cases} \epc
\end{equation}
where $p = |\a^+|$. Defining
\begin{equation}
     \ell_j = \begin{cases}
                 \e_j^+ & j = 1, \dots, p \\
                 \e_{m - j + 1}^- & j = p + 1, \dots, m
              \end{cases} \epc
\end{equation}
we obtain $x_j = m - \ell_j + 1$, $j = 1, \dots, m$. Finally, setting
$F_{\ell_j}^+ (\m) = F_j (\m)$ and $F_{\ell_j}^- (\m) =
\overline{F}_j (\m)$ and using (\ref{densdens}), we arrive at 
(\ref{multint}).

\section{Relation with previous results}
\label{app:alpha}
In our previous work \cite{BGKS07,BDGKSW08}, before we knew the multiple
integral representation for finite $\a$, we conjectured formulae which
we claimed to hold in the limit $\a \rightarrow 0$, relevant for the
physical correlation functions of the XXZ chain. We introduced a 
function, say, $\om_{\rm old} (\m_1, \m_2; \a)$  defined by an integral
formula involving, among other functions, a function $G_{\rm old}$
which is different from $G$ defined in (\ref{newg}). Here we explain
why our previous results remain unaltered in the limit $\a \rightarrow 0$
if we replace $\om_{\rm old} (\m_1, \m_2; \a)$ with
$- \om (\x_1, \x_2|\k, \a) + \om_0 (\x|\a)$ (the minus sign is due to a
change of conventions in which we followed \cite{BJMST08app,JMS08pp}).

Our previous ad hoc definitions were
\begin{align}
     & \om_{\rm old} (\n_1, \n_2; \a) - \om_0 (\x|\a) =
        - \x^\a \Ps_{\rm old} (\n_2, \n_1; - \a) + \D \ps(\x) \epc \\[1ex]
     & \Ps_{\rm old} (\n_2, \n_1; - \a) =
        2 \int_{\cal C} \rd m_0 (\m) G_{\rm old} (\m, \n_2; - \a)
          \bigl(q^\a \cth(\m - \n_1 - \h) - \cth(\m - \n_1) \bigr) \epc
          \notag
\end{align}
where $\rd m_0 (\m) = \rd m (\m)|_{\a = 0}$ and where $G_{\rm old}$ was
the solution of the linear integral equation
\begin{multline} \label{oldg}
     G_{\rm old} (\la, \n; - \a) = \\
         q^{-\a} \cth(\la - \n - \h) - \cth (\la - \n) 
         + \int_{\cal C} \rd m_0 (\m) K_\a (\la - \m) G(\m, \n; - \a) \epp
\end{multline}

Comparing with (\ref{defom}), (\ref{Psi}) and (\ref{newg}) we see that
apart from some conventional sign changes the only difference is that
the function $\r$ is replaced by unity in the old definitions. Using
that $\om_0 (\x|0) = 0$ we conclude in particular that
\begin{equation} \label{omomold}
     \om(\x_1, \x_2|\k, 0) = - \om_{\rm old} (\n_1, \n_2; 0)
\end{equation}
and that $G(\la, \n)|_{\a = 0} = G_{\rm old} (\la, \n; 0) = G_0 (\la, \n)$
which satisfies the integral equation
\begin{equation} \label{gnull}
     G_0 (\la, \n) = \re (\n - \la)
        + \int_{\cal C} \rd m_0 (\m) K(\la - \m) G_0 (\m, \n) \epp
\end{equation}
This integral equation implies the symmetry
\begin{equation} \label{symom}
     \om(\x_1, \x_2|\k, 0) = \om(\x_2, \x_1|\k, 0) \epp
\end{equation}

Let us define
\begin{subequations}
\begin{align}
     \om'(\x_1, \x_2) & = \6_\a \bigl(
        \x^{-\a} \om(\x_1, \x_2|\k, \a) \bigr) \bigr|_{\a = 0} \epc \\[1ex]
     \om'_{\rm old} (\n_1, \n_2) & = \6_\a \bigl(
        \x^{-\a} \om_{\rm old} (\n_1, \n_2; \a) \bigr) \bigr|_{\a = 0}
        \epp
\end{align}
\end{subequations}
We shall show below that
\begin{equation} \label{asymom}
     \om'_{\rm old} (\n_1, \n_2) =
        \tst{\2} \bigl(\om'(\x_2, \x_1) - \om'(\x_1, \x_2) \bigr) \epp
\end{equation}
Accepting this for a moment let us insert $- \om + \om_0$ instead of
$\om_{\rm old}$ into our previous formula for the exponential form,
e.g.\ into equation (35) of \cite{BDGKSW08}, which actually means
to use (\ref{Omega1}) for the $\tv$ independent part of the exponential
form. Then using (\ref{omomold}), (\ref{symom}) and (\ref{asymom}) and
the fact that $\x^{-\a} \om_0 (\x|\a) = {\cal O} (\a^2)$ we recover our
previous result, namely equation (37) of \cite{BDGKSW08}, in the limit
$\a \rightarrow 0$.

It remains to prove (\ref{asymom}). For this purpose consider
\begin{multline} \label{idapb1}
     \tst{\2} \bigl(\om'_{\rm old} (\n_1, \n_2) + \om'(\x_1, \x_2) \bigr)
        = \bigl(\r'(\x_1) - \r'(\x_2) \bigr) \ps(\x) \\
        - \r'(\x_1) \int_{\cal C}
                       \rd m_0 (\m) G_0 (\m, \n_2) \cth(\m - \n_1)
        - \int_{\cal C} \rd m_0 (\m) \r'(\re^\m)
          G_0 (\m, \n_2) \re(\n_1 - \m)
        \\
        + \int_{\cal C} \rd m_0 (\m) G' (\m, \n_2) \re(\n_1 - \m) \epc
\end{multline}
where $\r' (\x) = \6_\a \r (\x)|_{\a = 0}$, and $G' (\m, \n_2) =
\6_\a (G (\m, \n_2) + G_{\rm old} (\la, \n; \a))|_{\a = 0}$. Taking the
$\a$-derivative of (\ref{newg}) and (\ref{oldg}) we find
\begin{multline} \label{gstrich}
     G'(\la, \n_2) = - \r'(\x_2) \cth(\la - \n_2)
        - \int_{\cal C} \rd m_0 (\m) \r'(\re^\m)
                        G_0 (\m, \n_2) K(\la - \m) \\
        - \int_{\cal C} \rd m_0 (\m) K(\la - \m) G' (\m, \n_2) \epp
\end{multline}
Using (\ref{gnull}) and (\ref{gstrich}) we can eliminate $G'$ from
(\ref{idapb1}) by means of the `dressed function trick'. We arrive at
\begin{multline} \label{idapb2}
     \tst{\2} \bigl(\om'_{\rm old} (\n_1, \n_2) + \om'(\x_1, \x_2) \bigr)
        =  \bigl(\r'(\x_1) - \r'(\x_2) \bigr) \ps(\x) \\
        - \r'(\x_1) \int_{\cal C}
                       \rd m_0 (\m) G_0 (\m, \n_2) \cth(\m - \n_1)
        - \r'(\x_2) \int_{\cal C}
                       \rd m_0 (\m) G_0 (\m, \n_1) \cth(\m - \n_2) \\
        - \int_{\cal C} \rd m_0 (\m) \r'(\re^\m) G_0 (\m, \n_1)
                        G_0 (\m, \n_2)
\end{multline}
which is obviously symmetric. Then (\ref{asymom}) follows if one takes
into account that $\om'_{\rm old} (\x_1, \x_2)$ is antisymmetric, which
was shown in \cite{BGKS07}.

Now let us discuss how the operators $\tb_j$ described in the section 7
are related to the operators $\mathbf{h}_j$ (see formulae (40)-(42) of 
\cite{BDGKSW08}) in the limit $\alpha\rightarrow 0$. As we mentioned
above the operators $\tb_j$ have a pole of first order when
$\alpha\rightarrow 0$. On the other hand, it follows from the formula
(\ref{Omega2viat}) that the density matrix depends only on the
combination $(\rho_j-1)\tb_j$. Since $\lim_{\alpha\rightarrow 0}\rho_j=1$,
$1-\rho_j=\cal O(\a)$, and one obtains for $(\rho_j-1)\tb_j$ a finite
result in the limit $\alpha\rightarrow 0$. So we actually need only the
residues of $\tb_j$. Let us define 
\begin{equation} \label{tb0}
     \tb_j^{(0)} = \lim_{\a\rightarrow 0} (1-q^{\a})\tb_j \epp
\end{equation}
Since for the moment we do not have an explicit formula for $\tb_j$ in
the general case, let us describe the relation with $\mathbf{h}_j$ again
only for the cases $m = 1, 2$  where we know the explicit result
(\ref{tb1n1}) and (\ref{tb1n2}), (\ref{tb2n2}). It is enough to consider
$j=1$. 

For $m=1$ 
\begin{equation}
     \tb_{1[1,1]}^{(0)}=-\mathbf{h}_{1[1,1]}=-\frac12\;I\otimes\s^z_1 \epp
\end{equation}
For $m=2$ we simply get from (\ref{tb1n2}) 
\begin{equation}
     \tb_{1[1,2]}^{(0)}=-\frac14\;
     I\otimes\biggl[\s^z_1- \frac{q-q^{-1}}{\xi_1/\xi_2-\xi_2/\xi_1}
     \cdot(\s_1^+\s_2^- - \s_1^-\s_2^+)\biggr]
\end{equation}
and 
\begin{multline}
     \tb_{1[1,2]}^{(0)}+\mathbf{h}_{1[1,2]} \\
     = -\frac14\; \frac{\xi_1/\xi_2+\xi_2/\xi_1}
                       {\xi_1/\xi_2-\xi_2/\xi_1}\s^z_2\otimes
       \biggl[\s^z_1\s^z_2- \frac{q+q^{-1}}{\xi_1/\xi_2+\xi_2/\xi_1}
       \cdot(\s_1^+\s_2^- + \s_1^-\s_2^+)\biggr] \epp
\end{multline}

It is interesting to note the following. In spite of the fact that the
operators $\mathbf{h}_j$ are fermionic and the $\tb_j$ are bosonic, the
limit of the $\a$-trace of the corresponding exponentials should coincide 
\begin{equation} \label{fermi-bose}
     \lim_{\a\rightarrow 0} \;\;\mathbf{tr}^{\a} \bigl\{
     \exp\bigl(\sum_j \log{\rho_j}\;\tb_j\bigr)
     -\exp\bigl(-\sum_j \varphi_j\mathbf{h}_j\bigr)\bigr\}
     \bigl( q^{2 \a S(0)} \mathcal{O} \bigr)=0
\end{equation}
where $\varphi_j=\varphi(\xi_j|\kappa,0)$. We do not have a complete
understanding of this relation for the moment, but we can follow how
it works for $m=1,2$. The case $m=1$ is trivial while the case
$m=2$ is more interesting. For instance, one could expect that
terms containing $\varphi_1 \varphi_2$ appear, but they do not. For
the fermionic formula this is a simple consequence of the anti-commutation
relations of the operators $\mathbf{h}_j$. The reason why do they not
appear for the bosonic formula is different, namely, because of the
relation $\tb_1^{(0)}\tb_2^{(0)}=0$, which can be checked. We plan
to study this point more carefully in the future.
}        


\end{document}